\RequirePackage{fix-cm}
\documentclass{svjour3}                     % onecolumn (standard format)
\smartqed  % flush right qed marks, e.g. at end of proof
\usepackage[authoryear,comma]{natbib}
\usepackage{geometry}
\usepackage{amssymb,amsmath,mathrsfs,amscd}

\usepackage{multirow}
\usepackage{array}
\usepackage{parskip}
\usepackage[varg]{txfonts}

 \usepackage{gensymb}
\usepackage{longtable}
\usepackage{graphicx}
\usepackage{mathtools}
\usepackage{booktabs}
\usepackage{tabu}
 \usepackage{setspace}
\usepackage{wrapfig}
\usepackage{subfig}
\usepackage{epsfig}
\usepackage{lscape}
\usepackage{braket}
\usepackage{hyperref}
\usepackage{graphicx, color}

\newcommand{\doi}{doi}

\newcommand{\nat}{Nature}
\newcommand{\aap}{AAp}
\newcommand{\icarus}{Icarus}
\newcommand{\aj}{AJ}
\newcommand{\iaucirc}{IAUCirc.}
\newcommand{\procspie}{Proc.SPIE}
\newcommand{\aaps}{AAPS}
\journalname{Experimental Astronomy Journal}

\begin{document}
\title{High precision radial velocities with GIANO spectra}
\author{I. Carleo 
	\and N. Sanna
	\and R. Gratton
	\and S. Benatti
	\and M. Bonavita
	\and E. Oliva
	\and L. Origlia 
	\and S. Desidera 
	\and R. Claudi
	\and E. Sissa
    }
\authorrunning{I. Carleo et al.}
\offprints{I. Carleo,  \\
   \email{ilaria.carleo@oapd.inaf.it} }
   
\institute{I. Carleo \at
              INAF - Astronomical Observatory of Padua, Vicolo dell'Osservatorio 5, I-35122, Padova, Italy\\
              University of Padua, Department of Physics and Astronomy "Galileo Galilei", Via Marzolo, 8
              35131 Padova \\
              \email{ilaria.carleo@oapd.inaf.it}            \\
           \and
           N. Sanna \and E. Oliva \at
              INAF - Astronomical Observatory of Arcetri,Largo Enrico Fermi, 5, 50125, Florence, Italy 
            \and 
            R. Gratton \and S. Benatti \and S. Desidera \and R. Claudi \and E. Sissa  \at 
            INAF - Astronomical Observatory of Padua, Vicolo dell'Osservatorio 5, I-35122, Padova, Italy 
            \and 
            M. Bonavita \at 
            Institute for Astronomy, The University of Edinburgh, Royal Observatory, Blackford Hill, Edinburgh, EH9 3HJ, U.K.
            \and L. Origlia \at 
            INAF - Astronomical Observatory of Bologna, Via Ranzani, 1, 40127 Bologna, Italy
}

\date{Received: date / Accepted: February 17, 2016}

\maketitle

\abstract{Radial velocities (RV) measured from near-infrared (NIR) spectra are a potentially  excellent tool to search for extrasolar planets around cool or active stars. High resolution infrared (IR) spectrographs now available are reaching the high precision of visible instruments, with a constant improvement over time. GIANO is an infrared echelle spectrograph at the Telescopio Nazionale Galileo (TNG) and it is a powerful tool to provide high resolution spectra for accurate RV measurements of exoplanets and for chemical and dynamical studies of stellar or extragalactic objects. No other high spectral resolution IR instrument has GIANO's capability to cover the entire NIR wavelength range (0.95-2.45~$\mu$m) in a single exposure.\\
In this paper we describe the ensemble of procedures that we have developed to measure high precision RVs on GIANO spectra acquired during the Science Verification (SV) run, using the telluric lines as wavelength reference. We used the Cross Correlation Function (CCF) method to determine the velocity for both the star and the telluric lines. For this purpose, we constructed two suitable digital masks that include about 2000 stellar lines, and a similar number of telluric lines.\\
The method is applied to various targets with different spectral type, from K2V to M8 stars. We reached different precisions mainly depending on the $H$-magnitudes: for $H\sim 5$\ we obtain an rms scatter of $\sim 10$~m\, s$^{-1}$, while for $H\sim 9$\ the standard deviation increases to $\sim 50\div 80$~m\, s$^{-1}$. The corresponding theoretical error expectations are $\sim$4 m\, s$^{-1}$ and 30 m\, s$^{-1}$, respectively. Finally we provide the RVs measured with our procedure for the targets observed during GIANO Science Verification.

\keywords{}

%**************************************************************
\section{Introduction}

The search for exoplanets has led to more than 1900 discoveries through various detection techniques. About one third of known exoplanets have been detected with high-precision stellar radial velocity measurements in the visible wavelength region. Twenty years after the seminal discovery of 51 Peg-b by \cite{1995IAUC.6251....1M}, the RV technique is still one of the most important ones to discover planetary systems, and RV measurements are required to confirm planetary candidates found by photometric surveys.

The most favourable targets for RV measurements are solar-type stars (F, G, and K spectral types). They are generally observed at visible wavelengths for several reasons \citep{2010ApJ...713..410B}: these stars are bright at wavelength shorter than 1${\mu}m$ (visible region) where the spectra are rich in deep and sharp spectral lines, so a good Doppler shift measurement is possible; spectrograph technology operating in the visible region is more advanced relative to instruments operating at other wavelengths.

The most accurate RV measurements have been made with HARPS \citep{2004A&A...423..385P, 2006Natur.441..305L}. With a precision below 1~m\, s$^{-1}$, this instrument could reach planets down to a few Earth masses with short period orbits. Most discoveries are giant gaseous planets, hot-Neptunes and Jupiters, of short periods (few days). A few planets with masses between 1 and 10 Earth masses (super-Earths) have been discovered. Such small objects may be detected in favourable cases (inactive star) if they are close to the star, so that they are expected to be very hot due to the strong stellar irradiance. However, detection of such small planets around solar-type stars requires several tens to a few hundreds of high-quality RV points.

In the last years less massive stars, M-dwarfs, have become more interesting targets for various reasons, one of these being that M-dwarfs are more likely to host rocky planetary companions \citep{2010ApJ...713..410B}. In order to find habitable planets in orbit around solar-type star, the RV technique has to achieve a precision of 0.1~m\, s$^{-1}$. This constraint is released searching around less massive stars, because the reflex motion of the host stars due to the gravitational pull of the exoplanet is higher and more easily detectable than in the case of more massive stars. Moreover very cool stars such as M-dwarfs are the most numerous stars in the Galaxy \citep{2006AAS...209.2404H} and these stars have closer-in habitable zones than higher-mass stars \citep{1993Icar..101..108K}. This makes finding such planets easier: the small separation and shorter periods make the amplitude of the variation of RV large and therefore the temporal stability of the instrument is less constraining. The main problem with M-dwarfs is that they are much fainter at optical wavelengths, because they have effective temperatures of 4000~K or less, and they emit most of their spectral energy at wavelengths longer than 1~$\mu m$, so they can be better observed in the near-infrared region.
We know that RV signals can be induced by surface inhomogeneities, for example stellar spots \citep{2001A&A...379..279Q}, so a planet discovery can be confused with a variation due to such effects. An advantage of radial velocities measured from NIR spectra is that the jitter related to activity is reduced relative to visible measurements, because in the NIR the contrast between stellar spots or plagues and the rest of the stellar disk is reduced. Provided that RV can be measured with enough accuracy from NIR spectra, a comparison between variations of RV measured in the optical and NIR can establish the origin of the RV variations in an unambiguous way. 
For all these reasons there is a raising interest for measuring high precision RVs from NIR spectra. Recent technological improvements allowed to build more precise spectrographs for this spectral region, even if they don't reach yet the sub-m\, s$^{-1}$ precision of optical instruments. 

There has been some previous work in this area. An interesting case concerned the very young star TW Hya. \cite{2008Natur.451...38S} announced the discovery of a giant planet orbiting this star. They analyzed high resolution optical spectroscopic observations and obtained significant periodic radial velocity variation. This result, together with the lack of correlation between the RV variation and the cross correlation function bisector (BIS) seemed to be a proof of the existence of a planet orbiting the star. \\
\cite{2008A&A...489L...9H} studied the same object analyzing new optical and infrared data. The optical data were acquired with the CORALIE spectrograph at 1.2m Euler Swiss telescope in La Silla, Chile. Each measurement has an accuracy of about $10 m s^{-1}$. These data were complemented by older RVs measurements of TW Hya obtained with HARPS spectrograph and the FEROS, finding a period of about $3.56$ days and confirming the possible presence of a planet. \\
To further test this hypotesis Hu\'elamo et al. observed TW Hya over six nights in the infrared range with CRIRES, the CRyogenic high-resolution InfraRed Echelle Spectrograph mounted on the VLT. To derive the RV from the spectra they used the cross correlation method: the spectra were correlated with a telluric mask, developed with the HITRAN database, and a stellar mask from PHOENIX models. They found a dependence of the optical RV amplitude with the used CCF mask; the infrared RV curve is almost flat with a scatter of $35 m s^{-1}$. These results are inconsistent with optical orbital solution, so they concluded that the RV signal found for TW Hya is rather caused by a cool spot modulated by stellar rotation. \\
\cite{2008A&A...491..929S} investigated the intrinsic short-term radial velocity stability of CRIRES. This analysis was made both with gas cell calibrated data and on-sky measurements using the absorption lines of the Earth's atmosphere as local rest frames of radial velocities. They obtained observations of MS Vel, a M2II bright giant, over 5 hours. For the telluric lines, they used a standard atmospheric model for the Paranal site and adopted a typical humidity for the time of observation. They constructed a synthetic spectrum using the FASCODE algorithm and HITRAN database for molecular transitions. The radial velocity of the telluric lines is constant down to a level of $~10 m s^{-1}$ with the remaining residuals (rms) of about $20 m s^{-1}$. They showed that the telluric lines imprinted on the spectra of the science target are not a limiting factor in measurements of radial velocity, rather they can be used in substitution of gas cell as the RV zero point reference.\\
\cite{2010A&A...511A..55F} improved their previous work afore described \citep{2008A&A...489L...9H}. They analyzed the data of the radial velocity standard star, HD 108309, and of TW Hya over a time span of roughly one week acquired with CRIRES. In this work they used atmospheric features as wavelength reference. The RV values for the standard and TW Hya are compatible, within error bars, with the previously published values. TW Hya RV variation in the IR is not compatible with its optical counterpart, so the best explanation for these observed RV variations is a stellar spot, confirming the previous analysis, but with a better precision of $5-10 m s^{-1}$.

\cite{2010ApJ...723..684B}, using the NIRSPEC spectrograph on the Keck II telescope, obtained about 600~RV measurements over a period of six years for a sample of 59 late-M and L dwarfs to detect unseen companions. They developed a technique for measuring NIR RVs that makes use of CH4 absorption features in the Earth's atmosphere as a simultaneous wavelength reference. For a bright, slowly rotating M-dwarf standard they estimated an RV precision of 50~m\, s$^{-1}$, and for slowly rotating L-dwarfs a precision of 200~m\, s$^{-1}$.
\cite{2012ApJ...749...16B} presented the results of a high-resolution NIR RV analysis of twenty young stars in the $\beta Pic$ and TW Hya Associations. These spectra were acquired with NIRSPEC instrument. The determination of RVs was made using telluric absorption features as a wavelength reference. For each observation they created a model based on the combination of a telluric spectrum and a synthetically generated stellar spectrum. Both spectra were convolved by a parametrized instrumental profile and projected on a parametrized wavelength solution. The RV of the star was determined by minimizing the $\chi^2$ of the fit. The RV precision achieved with this method was of 50~m\, s$^{-1}$ for old field mid-M dwarfs. The observed RV dispersions for young stars were between 48~m\, s$^{-1}$ and 197~m\, s$^{-1}$. These dispersions were affected by noise from stellar activity or stellar jitter. The contribution of this effect was determined by subtracting, in quadrature, the average instrumental noise of 46~m\, s$^{-1}$, and the calculated theoretical noise, equal to 40~m\, s$^{-1}$, from the observed dispersions. The dependence of stellar jitter with projected rotational velocity limited the precision of 77~m\, s$^{-1}$ for the slowest rotating stars, 108~m\, s$^{-1}$ for modest rotating stars, and 168~m\, s$^{-1}$ for rapidly rotating stars. As expected, the NIR RV measurements decreased the RV noise caused by star spots by a factor of 3 compared to optical measurements.\\
One of the limiting factors of all these studies was the use of telluric lines to create a reference system. While this does not require any hardware investment, telluric lines are likely not stable enough to provide a reference system for precision well below 10~m\, s$^{-1}$. \cite{2010ApJ...713..410B} experimented an ammonia cell to create a suitable reference on the ESO CRIRES spectrograph. They obtained a precisions of $\sim 5$~m\, s$^{-1}$ over a six-month timescale and precisions of better than 3~m\, s$^{-1}$ over a timescale of a week. However, this cell was later dismounted and high precision RV from NIR spectra remain scarce. 
 
\cite{2010A&A...515A.106F} investigated the stability of atmospheric lines over long time-scales and at different atmospheric and observing conditions in order to quantify the precision of this kind of wavelength reference. Using HARPS data spanning 6 years and a telluric mask composed only of $O_{2}$ lines built from HITRAN database, they measured radial velocity variations for three bright stars (Tau Ceti, $\mu$ Arae and $\epsilon$ Eri) and obtained a long-term stability of telluric lines of 10~m\, s$^{-1}$, and a short-time-scales stability of 5~m\, s$^{-1}$, which goes down to 2~m\, s$^{-1}$ by using an atmospheric model to take into account the atmospheric phenomena.\\

In this paper, we describe in detail the ensemble of IDL (Interactive Data Language, a programming language used for data analysis) procedures created in order to measure RVs on near-infrared spectra acquired with GIANO spectrograph at TNG during the SV run. Since there is not yet any cell to be used with GIANO, our measurements used telluric lines as reference; however, our method can easily adapted with absorbing cells that are now planned in an upgrade of this instrument. We present the application of this technique for each target we have analysed, explaining the observed residuals and results that we obtained.

%**************************************************************
\section{Observations and data reduction}
\label{sec:data}

The data presented here were obtained with GIANO spectrograph in September 2014 during the SV run. This instrument is part of the Second Generation Instrumentation Plan of the Telescopio Nazionale Galileo (TNG) located at Roque de Los Muchachos Observatory (ORM), La Palma, Spain. GIANO is a cryogenic infrared cross-dispersed echelle spectrograph, which can yield, in a single exposure, 0.95-2.45~$\mu$m spectra at a resolution R${\sim}$50,000 \citep{2006SPIE.6269E..19O}. The dispersing element is a commercial 23.2 ll/mm R2 echelle working at a fixed position in a quasi-Littrow configuration with an off-axis angle along the slit of a $\sim$5 degrees \citep{2012SPIE.8446E..3TO}. Cross dispersion is provided by a combination of two prisms used in double-pass. The detector is a 2048$\times$2048 pixel Hawaii-2 PACE array by Teledyne, allowing to image almost the whole spectral range over 40 orders, with only small missing regions at the longest wavelengths. 
GIANO was designed and built for direct light feed from the telescope. Unfortunately, the focal station originally foreseen was not made available when GIANO was commissioned. Therefore the spectrograph had to be placed on the rotating building and complex light-feed system had to be developed using a pair of IR-transmitting ZBLAN fibres with two separate opto-mechanical interfaces. The first interface is positioned at the telescope focus and is used to feed the light into the fibres; it also includes the guiding camera and the calibration unit. The second interface re-images the light from the fibres onto an image slicer and then feeds the cryogenic slit \citep[see][for more details]{2014SPIE.9147E..9NT}. Each fiber has a core diameter of 84 micron (1 arcsec on sky) and the distance between centers is 250 micron (3 arcsec on sky). During the observation, one fiber looks at the sky, and the other one at the target. Similarly, during the calibration one fiber looks at the sky and the other looks at the calibration lamp. The opto-mechanical interface is at room temperature and its position relative to the cryogenic spectrometer cannot be stabilized with the accuracy required by high precision RV measurements, producing a variable illumination of the slit. In order to limit the impact of this issue, we need to record a reference absorption spectrum simultaneously with the object. Since absorption cells cannot be mounted in the current interface we can only use the telluric lines for this purpose.
It should be noticed however that while the use of absorbing reference lines may improve accuracy of RVs by nearly two orders of magnitude, some residual error may still be present, mainly because the profiles of stellar and telluric lines are intrinsically different. Such residuals are expected to depend e.g. on line strength.

All spectra of the stars were acquired with the {\it nodding-on-fiber} technique: target and sky were taken in pairs and alternatively acquired on fiber A and B (AB cycles), respectively, for an optimal subtraction of the detector noise and background. Calibration lamp (flat and U-Ne) spectra were acquired in day-time in stare mode, since lamps are diffuse sources and illuminate both fibers simultaneously. Typically, data were acquired with exposure times of 5 minutes.

Both standard RV and standard telluric stars were observed. The latter were chosen from a catalogue of several possible standard stars in the HARPS-N catalogue, in order to be visible during the observing nights, and bright enough to obtain a good $S/N$ in relatively short integration times. In particular for standard telluric stars, very hot rapidly rotating stars like B and O-type dwarfs are often chosen. In fact the small number of strong stellar lines makes them a good approximation of a continuum source suitable to observe the telluric spectrum. On the contrary RV standard stars require a much higher number of lines, and are therefore chosen among later spectral types.

The data were reduced and calibrated in wavelength as follows. To extract and wavelength-calibrate the GIANO spectra, the ECHELLE package in IRAF was used with some new \textit{ad hoc} scripts that are grouped in a package named GIANO\_TOOLS, a public library aimed to reduce and extract GIANO calibrated spectra by using routines available in any basic installation of IRAF (http://www.tng.iac.es/instruments/giano/). 2D$-$spectra of halogen lamps were used to identify the 49 orders of the echellogram and to map the geometry of the four spectra (two per fiber due to the slicer) in each order, for optimal extraction purposes. These solutions were applied to all the (A-B) 2D$-$spectrum computed from each pair of target exposures. The four spectra in each order were independently extracted and wavelength-calibrated, to minimize feature smearing in wavelength due to the small distortion of the slit image along the spatial direction induced by the off-plane illumination of the grating. The instrument is stable enough that flat-fields taken during the day-time are perfectly suited for this purpose. Each extracted spectrum is wavelength-calibrated by using the U-Ne lamp reference spectra taken at the beginning and/or at the end of the night. We used a set of $~$30 bright lines (mostly Ne lines) distributed over a few orders to obtain a first fit, then the optimal wavelength solution is computed by using ~300 U-Ne lines distributed over all orders. This method allowed us to reach a high accuracy of the wavelength calibration \citep{2014SPIE.9147E..1EO}.

The observed targets with corresponding characteristic (taken from the astronomical databases \footnote{http://simbad.u-strasbg.fr/simbad/sim-fid}) are listed in Table~\ref{tab:targets}.

%\begin{landscape}
\begin{table*}[tb]\footnotesize
\centering
\caption{Available data from the Science Verification run.}
\begin{tabular}{lllllccccc}

\toprule
Name   & R.A. J2000 &  Decl. J2000   &  Sp.Type & Classification &  I & J & H & K & RV  \\
   &   &     &   &  &   &  &  &  &  (\emph{km\, s$^{-1}$}) \\
\midrule
HD3765    & 00 40 49.269 & +40 11 13.83 & K2V   & RV-standard & 6.30  &  5.69 & 5.27 & 5.16 & -63.32\\
GJ1214    & 17 15 18.94  & +04 57 49.7 & M4.5V &	Planetary transit  & 11.10 & 9.75  & 9.09 & 8.78 & 21.10 \\
Gl15A     & 00 18 22.885 & +44 01 22.628 & M2.0V & Flare stars & 6.40  & 5.25 & 4.48  & 4.02  & 11.62 \\
VB10      & 19 16 57.622 & +05 09 02.18 & M8.0V & Variable star  & -  & 9.91  & 9.23 & 8.77 & -35.50 \\
HIP029216 & 06 09 39.574 & +20 29 15.458 & O6V   & Telluric   & -    & 7.31  & 7.37 & 7.39 & 23.20 \\
HIP89584  & 18 16 49.655 & -16 31 04.313 & O6.5V & Telluric   & -    & 7.45  & 7.39 & 7.35 & -53.00 \\
\bottomrule
\end{tabular}
\label{tab:targets}
\end{table*}

%\begin{table}[tb]
%\caption{Number of spectra for each science target.}
%\centering
%\begin{tabular}{llc}
%\toprule
%Name	  &  Julian Day   & Number of spectra\\
%\midrule
%\multirow{2}*{HD3765} & 2456907.5 &  4 \\
%                      & 2456910.5 &  3 \\
%   				  & 2456911.5 &  4 \\
%     				  & 2456912.5 &  3 \\
%   				      & 2456913.5 &  3 \\
%  				      & 2456914.5 &  3 \\
% 				      & 2456916.5 &  3 \\
%\midrule
%\multirow{2}*{GJ1214} & 2456910.5 &  4 \\
%					  & 2456911.5 &  4 \\ 
%					  & 2456912.5 &  4 \\
%				      & 2456913.5 &  4 \\
%					  & 2456914.5 &  4 \\
%\midrule
%\multirow{2}*{Gl15A} & 2456907.5 &  4 \\
%					& 2456911.5 &  4 \\
%					& 2456912.5 &  3 \\
%					& 2456913.5 &  3 \\
%					& 2456914.5 &  3 \\
%					& 2456916.5 &  3 \\
%\midrule
%\multirow{2}*{VB10} & 2456911.5 &  5 \\
%					& 2456912.5 &  4 \\
%					& 2456913.5 &  4 \\
%					& 2456914.5 &  4 \\
%\bottomrule
%\end{tabular}
%\label{tab:data}
%\end{table}

%**************************************************************

\section{Sample analysis} 
\label{sec:sampleanalysis}

An ensemble of IDL procedures was created to measure RVs with the Cross Correlation Function (CCF) method \citep{1979VA.....23..279B}. The various steps of our procedure are detailed in this Section.

%**************************************************************
\subsection{Pre-reduction: spectrum normalization}

First of all, we re-sampled all original spectra to have a constant step in RV. This step is required to have uniform wavelength scales for all spectra. With the adopted method (CCF), a dense sampling allows to reduce RV errors because the wavelengths of the spectral lines are determined with an error that is at best equal to half this step due to sampling of the mask used in the procedure.  However, execution of the procedure becomes slow when large files are used. The re-sampling should be made with the same step for stellar and telluric lines. In order to choose an appropriate step some tests were performed, applying to the analysis different step values (starting from 800 and going down until 100 m/s). The best compromise between accuracy and computation time is a step of 200~ m\, $s^{-1}$ that is small enough that uncertainties in the wavelength of the lines in the mask do not introduce avoidable noise, but large enough to keep time required for execution of the procedure reasonable (a few minutes per spectrum). Each individual order of the input spectrum was then re-sampled to the new grid, using a third degree cubic spline interpolation, and normalized to an approximate continuum. The normalization spectrum is obtained by dividing the spectra for a fiducial continuum. This was obtained either considering a cubic spline interpolation through local maxima within specified spectral window (for high $S/N$\ spectra) or simply heavily smoothing the original spectrum (for low $S/N$\ spectra). In order to reduce the impact of cosmic rays and bad pixels, all spectral points with intensity normalized to a fiducial continuum larger than $1.2$ were set at 1.2.

%**************************************************************
\subsection{Subtraction of telluric contribution}
\label{subsec:Subtell}

In order to obtain the stellar spectrum cleaned from the telluric lines, we create a median spectrum of the Earth atmosphere (hereinafter, telluric spectrum), which is subtracted from the normalized stellar spectra. This operation is repeated for each observation night. As the stellar and telluric spectra could have different airmass, the median telluric spectra above-mentioned was created taking into account the airmass by means of a cosecant law, as the stellar and telluric spectra could have different airmass. 
In the same way we obtained the telluric spectrum cleaned from the stellar contribution (see Section \ref{subsec:Substar}). This subtraction yielded better results than a division of the spectra because a division leads to large errors in correspondence of strong telluric lines. 
Therefore, for each science observation we obtained two cleaned spectra: the stellar spectra without telluric lines, and the telluric spectra without the stellar contribution. These spectra are used for the derivation of the stellar RV and of the rest RV corresponding to the telluric spectrum.

%**************************************************************
\subsection{Masks preparation}

%**************************************************************
\subsubsection{Stellar Mask}
\label{subsubsec:Smask}

The CCF method is performed by cross-correlating the spectrum with a mask. This is a vector with dimension equal to the observed spectrum, whose components are all zero, except those for which the condition $|\lambda_{\rm spectrum}-\lambda_{\rm line,i}|<step$, is satisfied, where $step=\lambda_1 - \lambda_0$, $\lambda_{\rm spectrum}$\ is the wavelength for the spectrum and $\lambda_{\rm line,i}$\ is the wavelength of the mask lines. In general the list of lines should include as many lines as possible in order to maximize the RV signal. To optimize the result, lines should be weighted accordingly to their strength on the spectrum. We need to prepare two masks, for the stellar and telluric spectra, respectively. In order to measure absolute RVs, the wavelength of the mask should be given by laboratory data; this requires identification of each individual line in the mask\footnote{While not strictly needed in our method, but in order to validate our procedure for the preparation of the mask, we actually counter-identified 757 of the 1102 lines of the mask we obtained for the K2V star HD~3765 with those listed in the solar spectrum tables of \cite{1953ApJ...118..397G}, \cite{1953ApJ...117...41M}, and in the NIST atomic spectra database (www.nist.gov/pml/data/asd.cfm). The average offset between measured and tabulated wavelengths is 0.0090 nm, with an r.m.s. scatter for individual lines of 0.0162~nm. Lines in our mask not counter-identified with solar spectrum lines are all weak, having a reduced equivalent width $\log{EW/\lambda}<-5.1$, where $EW$ is the equivalent width. Since HD~3765 is much cooler than the Sun, most spectral lines due to metals and molecules are stronger in its spectrum than in the Solar one. It is then not surprising that many lines that are weak in the spectrum of HD~3765 were not detectable in the Solar spectrum.}. However, we are interested here in variations of RVs rather than in their absolute values. In this case, what is important is that the same mask is used for all the spectra of a star, but the mask wavelengths do not need to be those observed at rest. A mask optimized for each individual star can be used.

We prepared an IDL procedure that automatically builds the list of lines and masks from the re-sampled, normalized and cleaned spectra (either stellar or telluric). The procedure works as follows. First, individual stellar spectra are shifted at rest velocity and their median is then obtained. For this purpose, the intrinsic RV of the star is taken from the astronomical databases SIMBAD, while the barycentric correction $BC$ is obtained through an IDL procedure previously prepared at Astronomical Observatory of Padua (OAPD). This obviously implies that the average RV measured with this mask will be that of the database. Previous tests showed that the procedure correctly gives the correction of RVs to the barycenter of the Solar System within a few hundredths of m\, s$^{-1}$, that is well enough for the present purposes.
The weights, i.e. the values for mask at the wavelengths of each line, are set at $F_{\rm center}$, that is the line intensity. This allows to weight the lines according to their intensity when computing the CCF. 
Only for the purpose of creation of the mask, the spectrum is changed by sign and summed 1, so that the absorption lines now appear as emission lines with a maximum intensity of 1. The line list is obtained by dividing each order into 128 chunks; each of them was searched for lines using the following method. For each chunk, the wavelength yielding the maximum flux was found. A short interval around it was considered, where the spectrum was fit by a four parameters Gaussian function: $a_0$ is the Gaussian height, namely the line intensity; $a_1$ is the center of Gaussian, that represents central wavelength of line; $a_2$ is the Gaussian width, which gives the Full Width at Half Maximum (FWHM) of the line and $a_3$ is the local continuum. Finally, the central intensity of line is given by $I_{\rm center}$ = $a_0/a_3$. In order to avoid blended lines, which are present in several orders, the software looks for pairs of lines whose separation is less than a critical value: for such pairs, only the line with the highest intensity is left in the mask list.

\begin{figure}[!ht]
    \centering
	\includegraphics[scale=0.5]{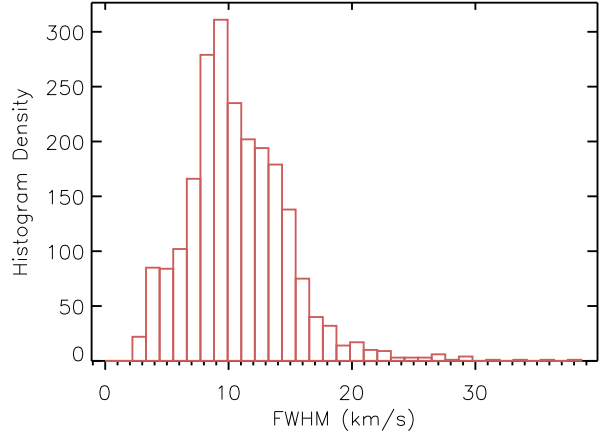}
	\caption{Density histogram of Full Width Half Maximum.}
	\label{fig:density histogram}
\end{figure}

To further clean the line list from artifacts, only the lines with a value of the FWHM in agreement with that expected for the rotational velocity of the star and the spectral resolution of the spectrograph should be considered. We then inspected the density histogram of the FWHM (Fig. \ref{fig:density histogram}) obtained for a slowly rotating star. This clearly shows that the value of this quantity peaks at $\sim 9$~km\, s$^{-1}$, in agreement with the expected value given the GIANO spectral resolution. Therefore only values of $0.5~R<FWHM<2.5~R (1+4a_0^2)$, with $R=\frac{\lambda_{\rm center}}{50000}=\frac{a_1}{50000}$\ were kept. While this is appropriate for slowly rotating stars, the most appropriate value of $R$\ shall in general be adopted considering the rotational velocity of the star.

%**************************************************************

\subsubsection{Telluric Mask}
\label{subsubsec:Tmask}

Just like for the stellar mask, the line list of the telluric target is obtained by considering the median telluric spectrum.  The telluric mask is built considering a line list that includes about 2000 lines, obtained with the normalized spectra of the telluric standard. As telluric lines have a different profile respect to stellar lines, the relative shifts between the fibers and the slit affect the two profiles in a different way and this can impact on the measure of RVs. In order to solve this inconvenient, only telluric lines with an intensity similar to the stellar ones are chosen.

%**************************************************************
\subsection{Subtraction of stellar contribution}
\label{subsec:Substar}

Not only the stellar spectra can be contaminated by the telluric contribution, but also there can be a contamination of  stellar lines in the telluric spectra. To avoid this issue, the normalized telluric spectra are subtracted from a stellar template, which is created by the median of stellar spectra (that was used for stellar mask, see Sec. \ref{subsubsec:Smask}) re-shifted by a factor $\lambda_{m}=\lambda_0(1+\frac{v_{geo}}{c})$, where $\lambda_{m}$ is the measured wavelength, $\lambda_0$ is the wavelength at rest, $v_{geo}$ = $v_{He} - BC$ is the geocentric velocity given by the difference between the heliocentric velocity and the barycentric correction, and $c$ is the speed of light.

%**************************************************************
\subsection{High precision RVs}

At this point, using an IDL procedure, RVs of both telluric and stellar lines are finally measured with the CCF method, including the following steps:

\begin{figure}[!ht]
	\centering
	\includegraphics[scale=0.5]{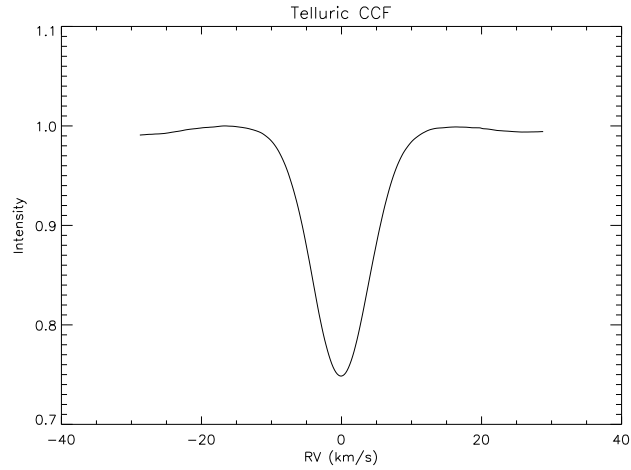}
	\caption{Total Cross Correlation Function of telluric spectrum.}
	\label{fig:ccf tell}
\end{figure}

\begin{figure}[!ht]
	\centering
	\includegraphics[scale=0.5]{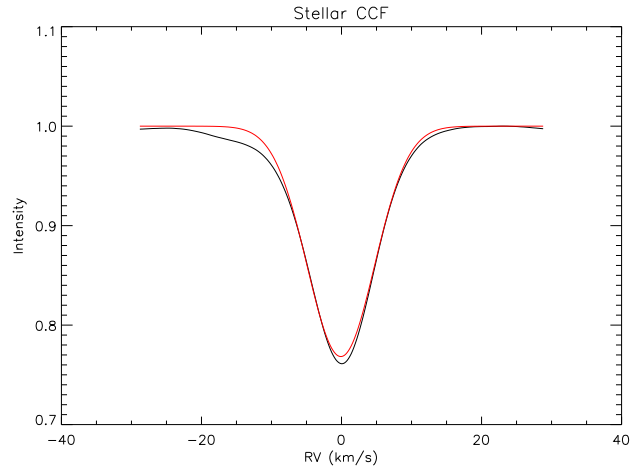}
	\caption{Total Cross Correlation Function of HD3765 star. The Gaussian fit is shown as a red line.}
	\label{fig:ccf star}
\end{figure}

\begin{enumerate}
\item \textbf{Reading input files}: the procedure works on the normalized stellar spectra subtracted by telluric contribution and on the normalized telluric spectra subtracted by stellar contamination, as input files.
\item \textbf{Telluric CCF for individual orders and error estimation}: In order to obtain the RV of the telluric lines, the telluric mask is cross-correlated with the normalized subtracted telluric spectra (Sec. \ref{subsec:Substar}) for each order and a Gaussian fit is subsequently executed for every CCF. The Gaussian fit provides four parameters which allow to obtain the FWHM and the RV. The centering error for each order is measured.
\item \textbf{Stellar CCF for individual orders and error estimation}: Likewise, in order to obtain the RVs of the star, the stellar mask is cross-correlated with normalized subtracted spectra of the star (Sec. \ref{subsec:Subtell}), but in this case, the intrinsic RV of the star has to be corrected by a factor $bb=\frac{v_{\rm barycentric}-BC}{step}$, where $v_{\rm barycentric}$ is the RV of the star with respect to the barycenter of the Solar System. The center of the line on the CCFs are then measured for each order obtaining the RVs with their errors, depending on FWHM, intensity and $S/N$.
\item \textbf{Weighted CCF}: The total weighted CCF for each spectrum is calculated for both the telluric (Fig. \ref{fig:ccf tell}) and the star (Fig. \ref{fig:ccf star}). Moreover the bisector is calculated for each spectrum (Fig.\ref{fig:bis_plot_star}).

\item \textbf{Order selection}: A number of spectral orders shows a paucity of lines in the corresponding masks, as well as poor $S/N$ due to a low atmospheric transmission at the relevant wavelengths. As expected, those orders show very large errors in the RV measurement, so we assign a null weight in the final solution. Since the accuracy is roughly proportional to the squared root of the number of orders, using only parts of an order would complicate the code and the calculation with a negligible gain, so we decide to fully reject or use an order. Basically we reject:
  \begin{itemize}
  \item[-] in the telluric spectra, all the orders with a measured RV that exceeds more than 1~km\, s$^{-1}$ in absolute value the value of the Earth's atmosphere lines, i.e. 0~km\, s$^{-1}$; 
  \item[-] in the standard RV stars spectra, all the orders with a measured RV that exceeds more than 1~km\, s$^{-1}$ the tabulated RV of the star.
  \end{itemize}	
\noindent A second selection is based on the value of the central intensity of the CCFs for the individual orders. The orders dominated by strong water vapour bands present very strong peaks in the telluric CCF and are located at the edges of each Y, J, H and K band. Few telluric lines are present at shorter wavelengths, making our methodology hard to apply in the Y band. For this reason we exclude it from our study. 
During the analysis we have found that, both for stellar and telluric lines, the best results are obtained when the central intensity of the CCF ranges between $0.2$ and $0.4$. The RV scatter is actually larger for orders with either weaker or stronger CCF intensity. In particular, for those orders with higher CCF intensity, the RV from stellar CCF is very unstable, probably due to the strong contamination by telluric lines. 
After the application of our selection criteria we can proceed with the analysis, using approximately half of the available orders. In these orders we still expect some spectrum-to-spectrum variations in radial velocities correlated with overall intensity of the telluric lines, due to the imperfect decontamination of the spectra.
\item \textbf{RV}: The RV for each spectrum was calculated by a Gaussian fit to the total CCF profile, providing the intensity, FWHM and RV values of the total CCF. The final RV for each spectrum is calculated by subtracting the telluric RV, $RV_{tell}$, from the stellar one $RV_{star}$: $RV=RV_{star}-RV_{tell}$, and the internal error is given by the final error, which takes into account the weight of each order.
\item \textbf{Internal Error Estimation}: The total error for the i-th order is given by the quadratic sum of both telluric and stellar contributions, and the final error for each spectrum, assumed that $S/N$ is only given by statistics of photons, is obtained weighting each order considered for the analysis. It is important to note that the internal error is calculated considering only photon statistics. There are also other noise sources, due to the S/N ratio differences caused by the different exposure times used, and by variations in the conditions of the Earth's atmosphere. There are also instrumental effects, e.g. temperature drift and differences due to the star itself, like variations in the stellar atmosphere due to magnetic activity, stellar oscillations, granulation, and so on. In addition, telluric lines may be not at rest with respect to the observer. Telluric lines are produced by a set of molecules (present at different heights) detectable at different wavelengths:
    \begin{itemize}
        \item $H_{2}O$ (ubiquitous): $<$5 km
        \item $O_{2}$ ($<$1.3 $\mu$m): $<$5 km
        \item $CO, CH_{4}$ (K-band): 10-20 km
        \item $CO_{2}$ (H-band): $>$30 km
    \end{itemize}
While in the visible range the strongest telluric lines are due to $O_2$ and $H_{2}O$ molecules in layers at low altitude, in the NIR also the molecules formed in layers at high heights are significant. For these lines we expect winds whose velocity component along the line of sight may well be as large as 10 m\, s$^{-1}$ that would reflect in offsets in the RVs. Moreover, since we re-sampled all spectra with a step of 200 m\, s$^{-1}$ in RV, each line has an associated wavelength error of half this step, i.e. 100 m\, s$^{-1}$. Since about 1000 lines are used for RV determinations, the resulting error for this source of noise is $100/\sqrt{N_{lines}}\sim3$~m\, s$^{-1}$, where $N_{lines}$ is the number of lines. This is not negligible, though it does not dominate the noise. 
\item \textbf{Corrections to RVs}: Finally, we applied a multivariate statistical analysis which allowed us to improve much more the results. By making this statistical analysis, we find correlations among various parameters, in particular between RVs and the Bisector Velocity Span (BVS) or intensity of the telluric spectrum, likely resulting from asymmetric slit illumination. As a result, the final RV is given after removing this correlation.  
\end{enumerate} 

\begin{figure}[!ht]
	\centering
	\includegraphics[scale=0.5]{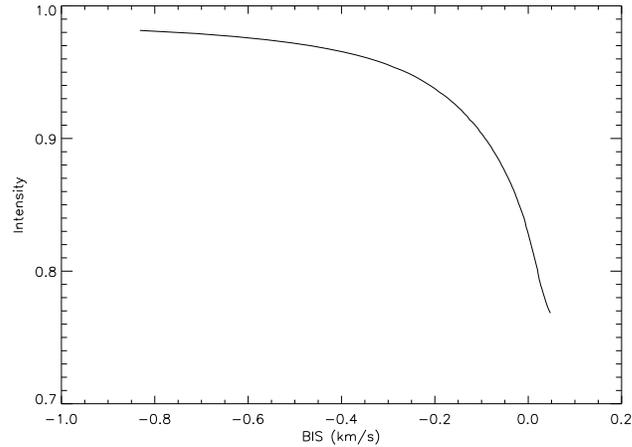}
	\caption{Bisector of HD3765.}
	\label{fig:bis_plot_star}
\end{figure}

\subsection{Bisector analysis}

A measure of the asymmetry of the bisector is given by the BVS. This is defined by comparing the position of the bisector at two flux levels of the profile of CCF (the top range lies around the 25\% of the flux and bottom range around the 75\%). The BVS is the difference between the median position of the bisectors in these two ranges.

The bisector analysis is important for two reasons:
\begin{itemize}
\item the RV of a star is defined to be the velocity of the center of mass of the star along our line of sight \citep{2001A&A...379..279Q}. The observational determination of a star's RV is made by measuring the Doppler shift of spectral lines. The RV variations can be  due to either a possible companion or changes in the stellar atmosphere. One of the best ways to interpret observed variations is to look for changes in the BVS of the stellar CCF. Any correlation between RV changes and line-bisector orientation leads to serious doubts on the reflex motion interpretation of the RV variations. If the RV is due to changes in center-of-mass velocity of the star, there is no change in bisector's shape or orientation.
\item In addition, in our analysis BVS significantly different from zero in the telluric CCF's may signal an asymmetric slit illumination, that may arise due to disalignments between the fiber ends and the slit. In this case, we might try to apply a correction based on the observed correlation between this quantity and the RVs. 
For all targets we performed a bisector analysis, studying the correlations between the stellar and telluric BVS through the Spearman coefficients. Significant correlations were found for GJ~1214 (r=0.47) and VB10 (r$\sim$0.05), while for HD~3765 there is not any correlation, but eliminating the second and the last observation night the significance increases with a Spearman coefficient of 0.74, indicating a strong correlation. These results suggest that an asymmetric slit illumination is frequent during the observations.
\end{itemize}

%**************************************************************
\section{Results}

\subsection{HD3765}

HD~3765 is a quite bright, solar metallicity \citep{2004A&A...418..551M}, K2V star with magnitudes J=5.69, H=5.27, and K=5.16 \citep{2003yCat.2246....0C}. It has a low activity level \citep{2000A&AS..142..275S,2010A&A...520A..79M} and a constant RV \citep[: RV jitter of 2.4 m\, s$^{-1}$]{1983BICDS..24....3B,2010A&A...524A..10C,2010ApJ...725..875I} and was then used as a standard for testing instrument performances. 

Twenty-three spectra of HD~3765 were acquired in seven nights, with typical S/N values of 130 and internal errors in RVs from 4.5 to 10 m\, s$^{-1}$. The original r.m.s. of the RVs we obtained was 28~m\, s$^{-1}$. 
Most of this scatter is probably due to mechanical vibrations and drifts between the warm-preslit system (that includes the fiber, the slicer and the re-imaging optics) and the cryogenic spectrometer \citep[see][for more details]{2014SPIE.9147E..9NT}. This was partly removed using the telluric line reference, that reduces the r.m.s. scatter of the stellar RVs from an original value of 610 to 28~m\, s$^{-1}$. However, even once corrected for the telluric reference, the RVs show significant correlations on the BVS of both the stellar and telluric CCFs (see Fig. \ref{fig:correl1_hd3765} and Fig.\ref{fig:correl2_hd3765}). Removing these correlations, by subtracting the best fit function from the RVs,  the r.m.s. decreases to 14~m\, s$^{-1}$ and 8~m\, s$^{-1}$ averaging observations taken in the same night. Table~\ref{tab:hd3765_0} summarizes the final result for each spectrum of HD~3765. The resulting RV values after corrections are plotted against the Julian Days in Fig.\ref{fig:hd3765_jd_rv}.

 \begin{figure}[!ht]
	\centering
	\includegraphics[scale=0.5]{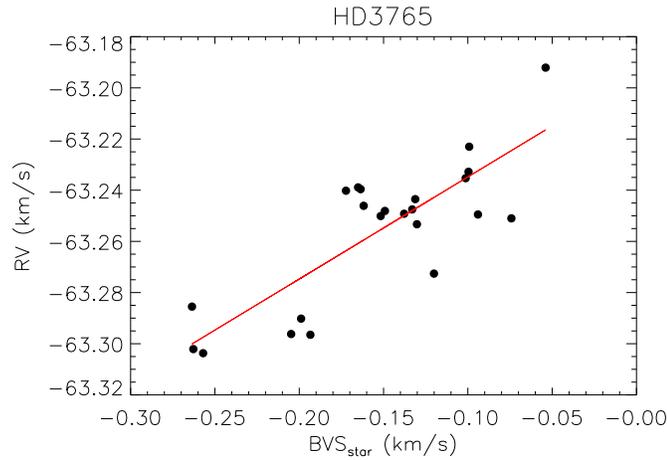}
	\caption{Correlation between RV values and the stellar bisector velocity span for HD3765. The best linear fit is represented by a red line}
	\label{fig:correl1_hd3765}
\end{figure}

 \begin{figure}[!ht]
	\centering
	\includegraphics[scale=0.5]{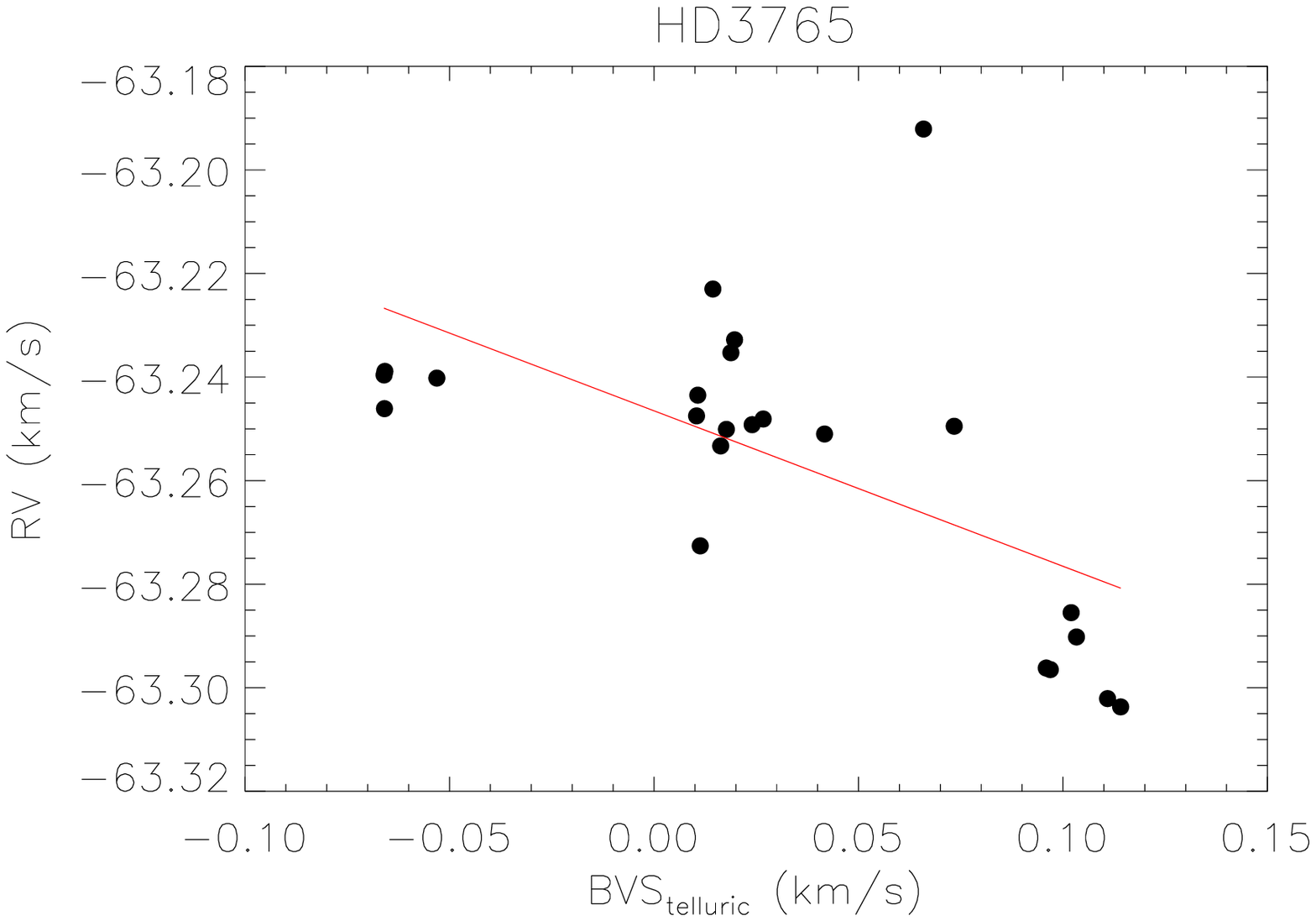}
	\caption{Correlation between RV values and the telluric bisector velocity span for HD3765. The best linear fit is represented by a red line.}
	\label{fig:correl2_hd3765}
\end{figure}

 \begin{figure}[!ht]
	\centering
	\includegraphics[scale=0.5]{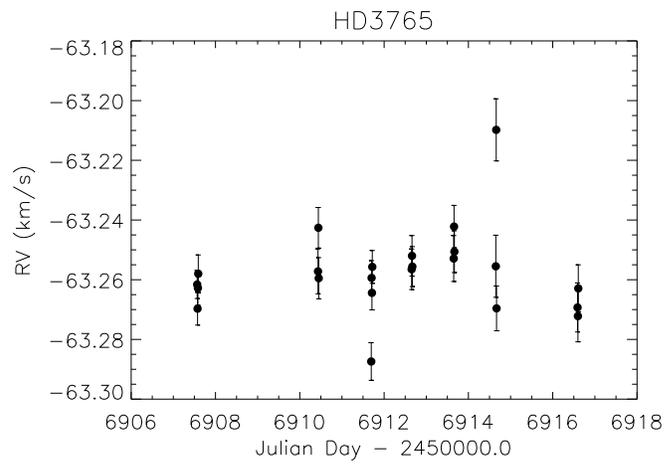}
	\caption{Radial velocity values against the Julian Days for HD3765.}
	\label{fig:hd3765_jd_rv}
\end{figure}

\subsection{GJ1214}

GJ1214 is an M4.5 red dwarf that hosts a transiting super-Earth planet, with a mass of 6.55 Earth masses and a period of 1.57 day \citep{2009Natur.462..891C}. The RV semi-amplitude of the orbit is about 13 m\, s$^{-1}$, which is challenging for GIANO. The star is very faint at optical wavelength (V=14.67), but easily observable with GIANO in the NIR: J=9.750, H=9.094, K=8.782 \citep{2003yCat.2246....0C}. The star is old (6 Gyr) and slowly rotating \citep{2011ApJ...736...12B}. We obtained 20 spectra over 5 nights, with a typical S/N of 25. We performed an analysis very similar to that considered for HD~3765, searching for signatures of the orbital motion. Due to the lower S/N of the spectra, the internal errors of the RVs are larger than for HD~3765, ranging from 29 to 45m\, s$^{-1}$. The r.m.s. scatter of the original RVs is 141 m\, s$^{-1}$, that reduces to 122 m\, s$^{-1}$ if we average results obtained at short cadence (less than 1 hr). This is clearly much larger than expected for the known RV curve and even for the internal errors we obtained for the GIANO RVs. In this case, there is a strong correlation of the residuals with respect to the known RV curve with the intensity of the telluric lines (Fig.\ref{fig:correl2_gj1214}). Once this is removed using a best fit line, the scatter of the RV is lowered to 111 m\, s$^{-1}$, that reduces to 65 m\, s$^{-1}$ average results from short cadence observations. This is the best we could obtain from the GIANO spectra, but still not enough to detect the orbital motion for this system (see Fig. \ref{fig:fase_gj1214}). Table~\ref{tab:gj1214_results} summarizes the final result for each spectrum of GJ~1214. The resulting RV values after corrections are plotted against the Julian Days in Fig.\ref{fig:GJ1214_jd_rv} .

\begin{figure}[!ht]
	\centering
	\includegraphics[scale=0.5]{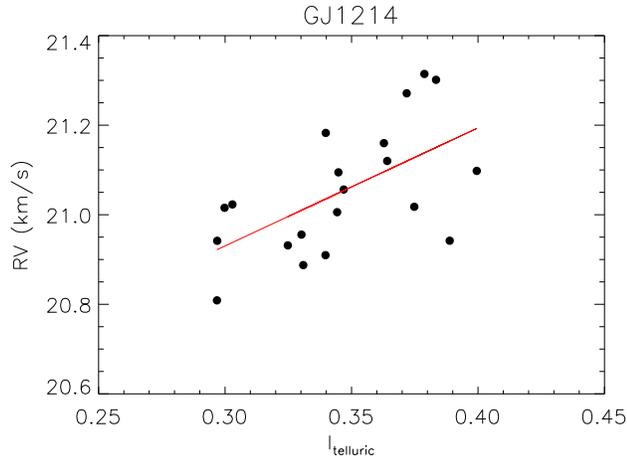}
	\caption{Correlation between RV values and the telluric line intensity for GJ1214. The best linear fit is represented by a red line.}
	\label{fig:correl2_gj1214}
\end{figure}

 \begin{figure}[!ht]
	\centering
	\includegraphics[scale=0.5]{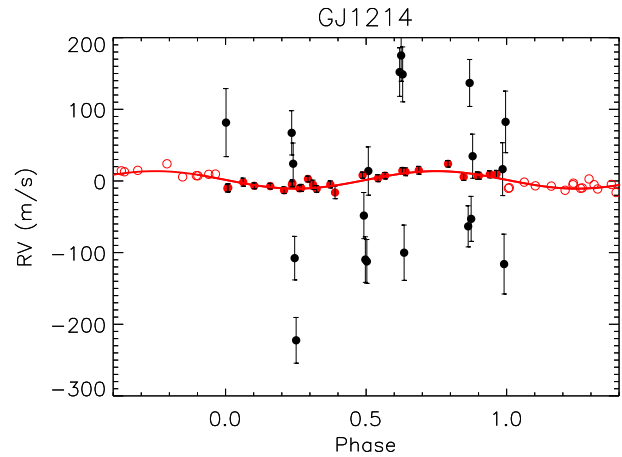}
	\caption{Radial velocity values against the orbital phase. The black points are the RV values obtained with GIANO; the red points are the RVs from \cite{2009Natur.462..891C} with corresponding fit.}
	\label{fig:fase_gj1214}
\end{figure}

 \begin{figure}[!ht]
	\centering
	\includegraphics[scale=0.5]{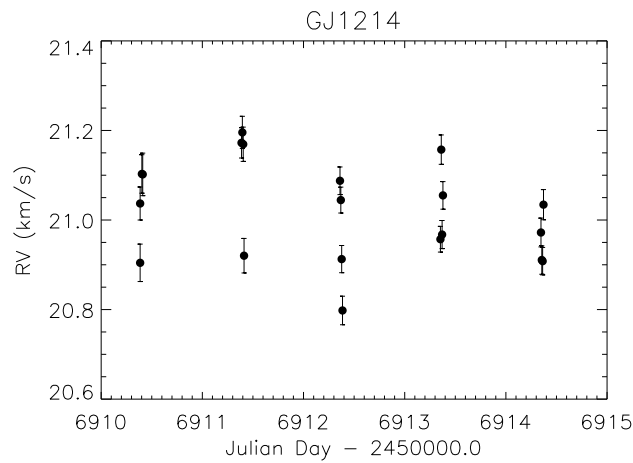}
	\caption{Radial velocity values against the Julian Days for GJ1214.}
	\label{fig:GJ1214_jd_rv}
\end{figure}

\subsection{Gl15A}

Gl15A is a bright, nearby M1.5 star (distance 9 pc) with magnitude J=5.25, H=4.48, and K=4.02 \citep{2003yCat.2246....0C}. As for GJ1214, a super-Earth planet, with a mass of $\sim 0.017$ Jupiter masses ($M_J$) and a period of about 11 days, has been discovered around this star \citep{2014ApJ...794...51H} from RV observation at visual wavelengths with the HIRES spectrograph at Keck. No transit has been observed and the amplitude of the RV curve (2.94 m\, s$^{-1}$) is clearly beyond the accuracy possible with GIANO. Hence, for the purposes of this study, we may consider this star as having a constant RV. We obtained 20 spectra over 6 nights, with typical S/N of 210. The internal errors in the RVs range from 2.7 to 5.6~m\, s$^{-1}$. The r.m.s. of the GIANO RVs is 34 m\, s$^{-1}$, with about the same value if we average observations obtained during the same night. As in the case of GJ1214, we found a strong correlation (Pearson correlation coefficient r=0.93) with the intensity of the telluric lines (Fig. \ref{fig:correl2_gl12a}). If we remove this strong trend, as above by subtracting the best fit function from the RVs, the r.m.s scatter of the RVs is reduced to 18 m\, s$^{-1}$, that is further reduced to 11 m\, s$^{-1}$ eliminating two outlier observations taken in the fifth night. Table~\ref{tab:gl15a_results} summarizes the final result for each spectrum of Gl15A. The resulting RV values after corrections are plotted against the Julian Days in Fig.\ref{fig:gl15a_jd_rv} .

 \begin{figure}[!ht]
	\centering
	\includegraphics[scale=0.5]{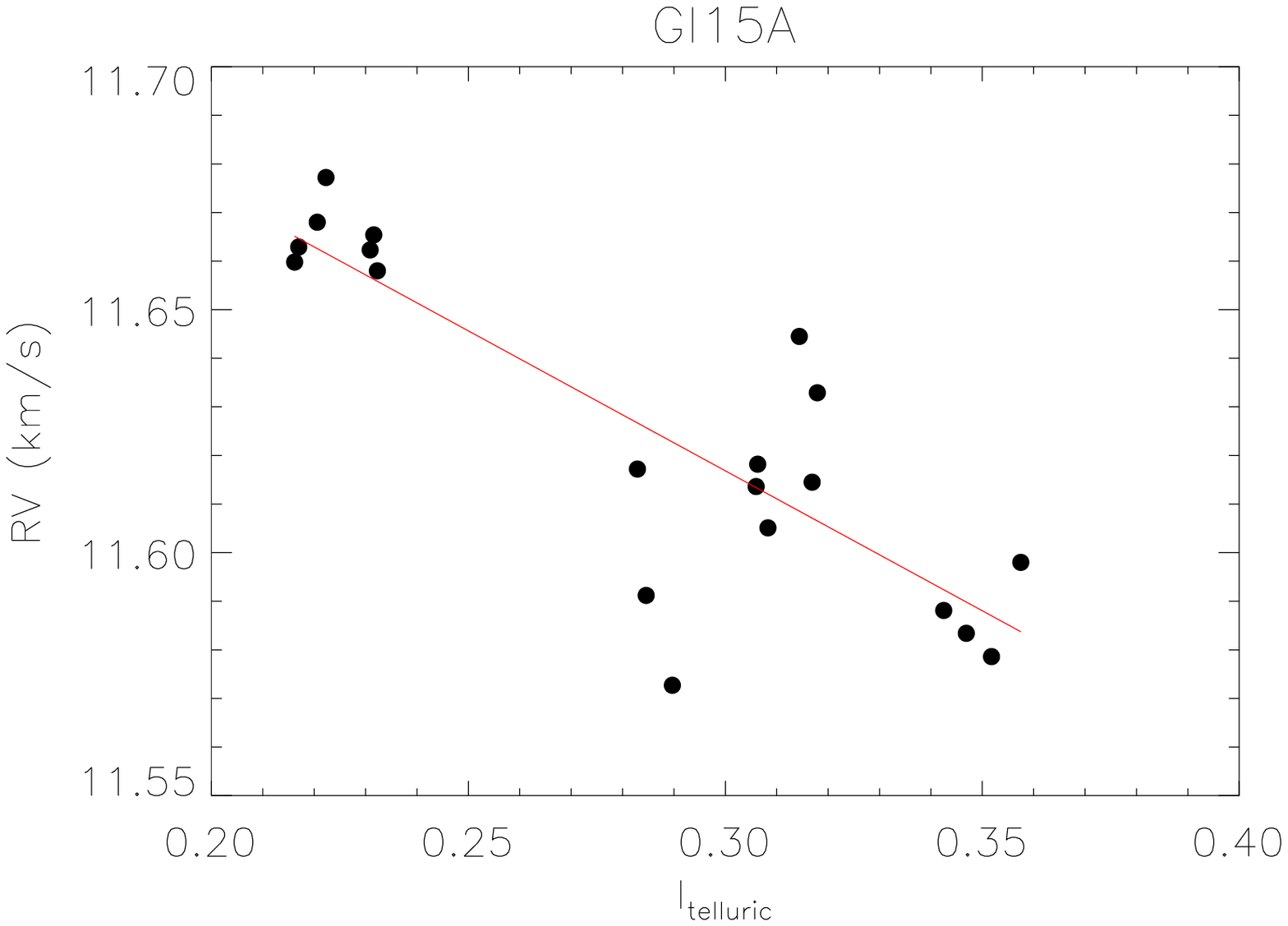}
	\caption{Correlation between RV values and the telluric line intensity for Gl15A. The best linear fit is represented by a red line.}
	\label{fig:correl2_gl12a}
\end{figure}

 \begin{figure}[!ht]
	\centering
	\includegraphics[scale=0.5]{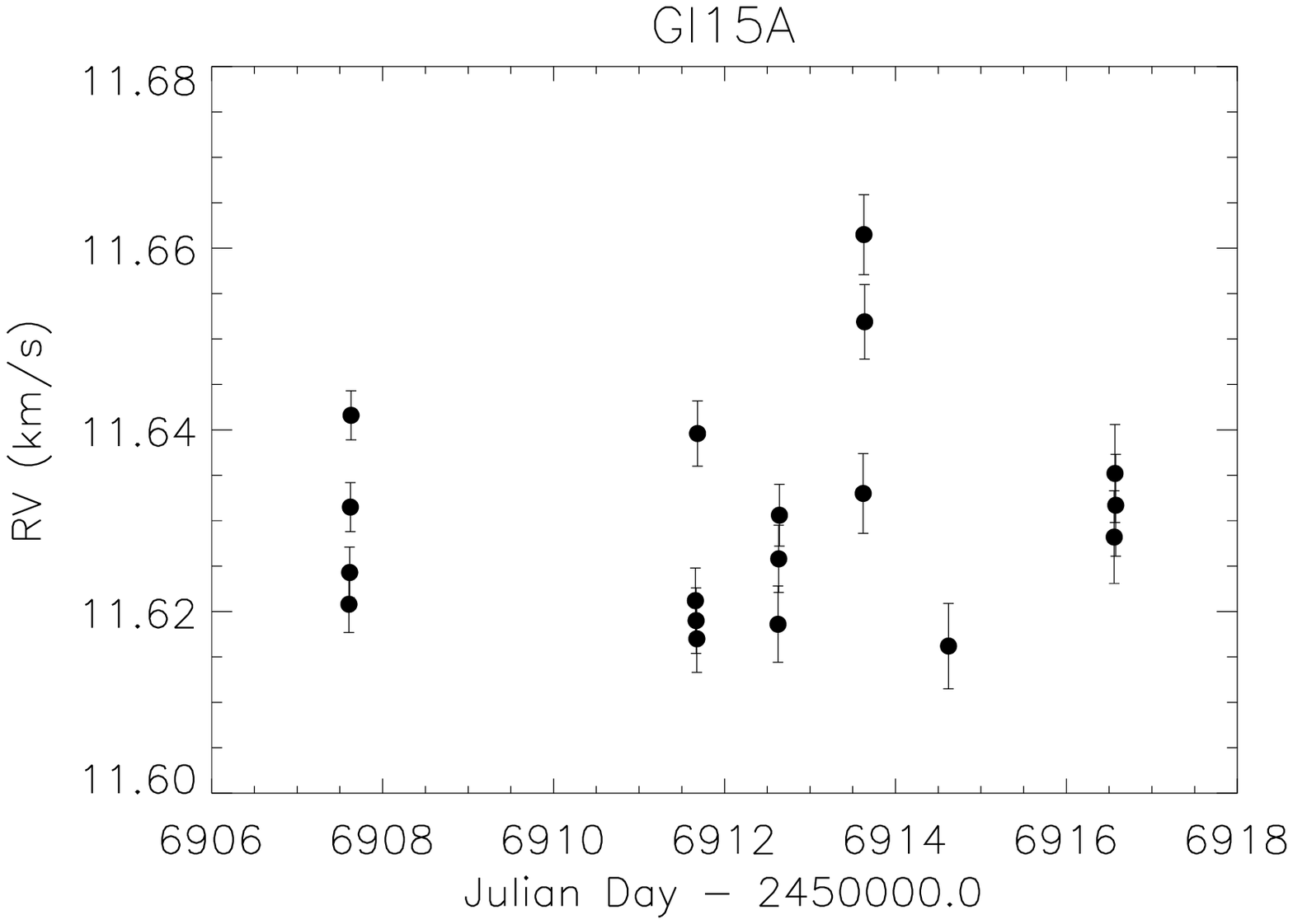}
	\caption{Radial velocity values against the Julian Days for Gl15A.}
	\label{fig:gl15a_jd_rv}
\end{figure}

\subsection{VB10}

The last star considered in this paper is VB10 (=GJ752B); this is a very close (distance 5.9 pc) M8.0V dwarf. The star has a moderate rotation ($V~\sin{i}=6.5$~km\, s$^{-1}$: \cite{2003ApJ...583..451M}). This star is very faint at optical wavelengths (V=17.30) making RV measurements at these wavelengths extremely difficult. It is much brighter in the NIR (J=9.908, H=9.226, K=8.765: \cite{2003yCat.2246....0C}); for this reason a few RVs (with errors larger than 200 m\, s$^{-1}$) have been obtained using NIRSPEC at Keck \citep{2012A&A...538A.141R}. \cite{2009ApJ...700..623P} announced  a  massive planet ($\sim 6.4$ $M_J$) around this star discovered by means of astrometrical data, with a period of $\sim 0.7$ yr; such a planet would induce RV variations as large as 1 km\, s$^{-1}$. However, this planet was later refuted by \cite{2010ApJ...711L..19B} using high-resolution spectra with CRIRES equipped with an Ammonia cell achieving an RV precision of $\sim 10$~m\, s$^{-1}$, and found no significant RV variability over about 7 months. For our purposes, we can again consider this star as having constant RV. We obtained 17 RVs over 5 nights; internal errors of individual measurements are between 30 and 40 m\, s$^{-1}$. The original r.m.s. of our measurements is 131 m\, s$^{-1}$ (eliminating as outlier the last observation of the first night), that reduces to 113 m\, s$^{-1}$ averaging observations taken in the same night. As for most of the other stars, also in this case we found a strong trend with intensity of the telluric lines  (Pearson correlation coefficient r=0.71, Fig. \ref{fig:correl2_vb10}); once removed, the scatter decrease to 92 m\, s$^{-1}$ for individual observations, and to 59 m\, s$^{-1}$ if we average results of the same night. While clearly much worse than the CRIRES results, this scatter is much lower than e.g. obtained with NIRSPEC at Keck \citep{2012A&A...538A.141R}. The two correlations in Fig. \ref{fig:correl2_gl12a} and Fig. \ref{fig:correl2_vb10} seem to be quite similar with a similar trend, but on a larger sample of stars one can see that this trend is random, also changing in slope, i.e. Fig. \ref{fig:correl2_gj1214}.\\
Table~\ref{tab:vb10_results} summarizes the final result for each spectrum of VB10. The resulting RV values after corrections are plotted against the Julian Days in Fig.\ref{fig:VB10_jd_rv} .

 \begin{figure}[!ht]
	\centering
	\includegraphics[scale=0.5]{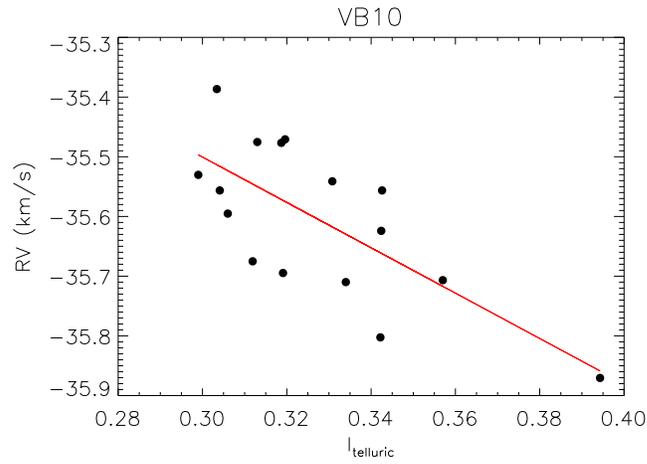}
	\caption{Correlation between RV values and the telluric line intensity for VB10. The best linear fit is represented by a red line.}
	\label{fig:correl2_vb10}
\end{figure}

 \begin{figure}[!ht]
	\centering
	\includegraphics[scale=0.5]{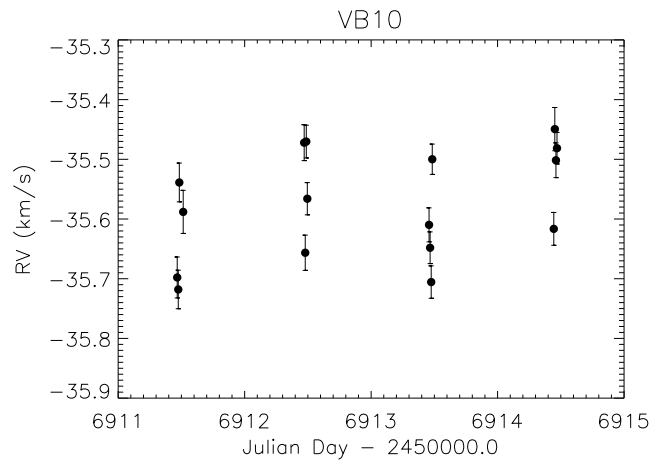}
	\caption{Radial velocity values against the Julian Days for VB10.}
	\label{fig:VB10_jd_rv}
\end{figure}

%**************************************************************
\section{Discussion and conclusions}

The goal of this work was to assess the precision of the RV measurements obtained using spectra acquired with the GIANO infra-red spectrograph, currently installed at the TNG in La Palma (Spain). The data used for this analysis were acquired during the September 2014 SV run. Since there was no other suitable set of reference lines available, we used the telluric lines as a zero point of the RVs. An ensemble of IDL procedures, developed specifically for this work, was used to perform all required steps in the analysis. We used the CCF method to determine the velocity for both the star and the telluric lines. To this purpose, we constructed two suitable digital masks that include about 2000 stellar lines, and a similar number of telluric lines. RV determinations include the following steps: preparation of files including evaluation of the correction to the barycenter of the solar system; normalization of spectra; cross correlation of individual orders with the appropriate masks (both stellar and telluric spectra) with derivation of individual CCF; weighted sum of the CCFs; derivation of RVs for both stellar and telluric spectra along with the internal errors; derivation of high precision RVs; derivation of the bisector of the CCF and of the bisector velocity span (for both stellar and telluric spectra). The whole procedure requires about 1.5 minutes per spectrum.

By analysing the different spectral types of stars we found a correlation between the H magnitude and the RV precision we reached: the smaller is the H magnitude the smaller is the error in measurements. Fig.  \ref{fig:type_err} shows a representation for the four targets of our sample: the stars with an H magnitude of about 5, have a precision of 10 m\, s$^{-1}$, while the stars with H magnitude of about 9 reach a precision of 60 - 70 m\, s$^{-1}$. High precision RV are then possible with GIANO using the telluric lines as reference, at least for stars with bright H-magnitude. 
The dispersion achieved with GIANO falls between NIRSPEC and CRIRES dispersions. The main reason is due to the different resolutions: NIRSPEC has a resolving power of 25.000, so stellar and telluric lines are not resolved and are dominated by the instrumental profile, while CRIRES has a resolving power of 100.000 and it is dominated by the intrinsic lines profile respect to the instrumental one.

Starting from these results, we can estimate the number of stars accessible to GIANO/TNG for a given radial velocity precision, that spans from 30.000 to 150.000 depending on the H magnitude, as shown in the Table \ref{tab:numberstars}.

\begin{table*}[tb]
\centering
\caption{Number of stars accessible to GIANO/TNG for a given RV precision (data based on the 2MASS all-sky catalog of point Sources \citep{2003yCat.2246....0C}).}
\begin{tabular}{crc}
\toprule
H magnitude & RV precision  & Number of stars \\
 &  (m\, s$^{-1}$) &  ($10^{4}$)   \\
\midrule
5.00  &  10 & 3 \\
5.75  &  20 & 4.6 \\
6.50  &  40 & 5.6 \\
7.50  & 100 & 15 \\
 \\
\bottomrule
\end{tabular}
\label{tab:numberstars}
\end{table*}

In order to estimate the number of dwarfs accessible to GIANO we consider the all-sky catalogue in \cite{2011AJ....142..138L}: they selected 8889 M dwarfs from the SUPERBLINK survey of stars with apparent infrared magnitude J$<$10 and spectral type from K7 to M7, and of these 655 (520 early-M, 16 late-M) have H$<$7.5 and are detectable with GIANO. Considering a planet with a mass of $10M_{\oplus}$ orbiting around an M2 dwarf, we calculate the habitable zone by entering the stellar physical characteristics \citep{2009ApJ...698..519K} in a tool online (http://depts.washington.edu/naivpl/sites/default/files/hz.shtml) and then we obtained the theoretical RV signal (K), which is about 3.65 m\, s$^{-1}$ for $i=90\degree$ with a $S/N\sim5$. Similarly, we calculate the RV signal for a $10M_{\oplus}$ planet orbiting an M5 dwarf, obtaining K=11.5 m\, s$^{-1}$. This means that one needs about 250 observations to detect an Earth-like planet for stars with H magnitude less than 6.5.\\

This paper has provided the precision reached with the current condition of GIANO. There is now a plan to improve the GIANO performances within the realization of a common feeding for GIANO and HARPS-N (GIARPS). For what concern GIANO this includes the elimination of the optical fibers and the insertion of a stable feeding through a train of optics, that will include a tip-tilt mirror located on an image of the telescope pupil and controlled in closed loop by a slit viewing camera. This train of optics will also allow insertion of an ammonia absorbing cell, similar to that used for CRIRES. With this new configuration of GIANO, the internal errors will be reduced and the number of observations required to detect a $10M_{\oplus}$ planet around an M5 dwarf above discussed could decrease down to $\sim$100. This is feasible in the equivalent of $\sim$10 observing nights.

 \begin{figure}[!ht]
	\centering
	\includegraphics[scale=0.5]{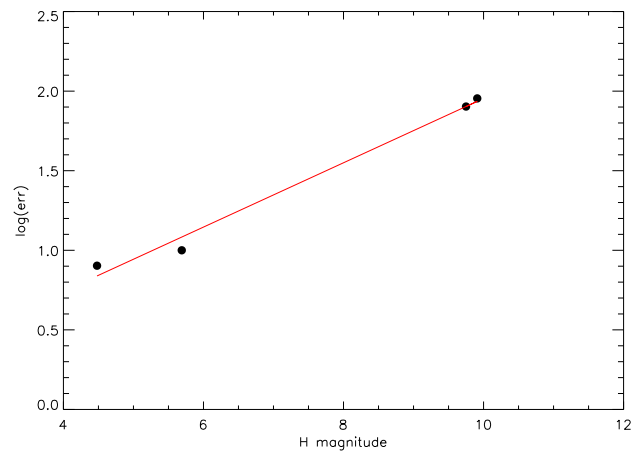}
	\caption{The error in the RV measurements depending on the H magnitude. The best linear fit is represented by a red line.}
	\label{fig:type_err}
\end{figure}

%**************************************************************

\begin{acknowledgements}
The authors acknowledges support from INAF through the "T-REX Progetto Premiale" funding scheme of the Italian Ministry of Education, University, and Research.
\end{acknowledgements}

\bibliographystyle{spbasic}

\begin{thebibliography}{}

\bibitem[{{Bailey} et~al(2012){Bailey}, {White}, {Blake}, {Charbonneau},
  {Barman}, {Tanner}, and {Torres}}]{2012ApJ...749...16B}
{Bailey} JI III, {White} RJ, {Blake} CH, {Charbonneau} D, {Barman} TS, {Tanner}
  AM, {Torres} G (2012) {Precise Infrared Radial Velocities from Keck/NIRSPEC
  and the Search for Young Planets}. ApJ 749:16,
  %\doi{10.1088/0004-637X/749/1/16}, %\eprint{1202.0300}

\bibitem[{{Baranne} et~al(1979){Baranne}, {Mayor}, and
  {Poncet}}]{1979VA.....23..279B}
{Baranne} A, {Mayor} M, {Poncet} JL (1979) {CORAVEL - A new tool for radial
  velocity measurements}. Vistas in Astronomy 23:279--316,
  %\doi{10.1016/0083-6656(79)90016-3}

\bibitem[{{Batten}(1983)}]{1983BICDS..24....3B}
{Batten} AH (1983) {Standard Velocity Stars}. Bulletin d'Information du Centre
  de Donnees Stellaires 24:3

\bibitem[{{Bean} et~al(2010{\natexlab{a}}){Bean}, {Seifahrt}, {Hartman},
  {Nilsson}, {Reiners}, {Dreizler}, {Henry}, and
  {Wiedemann}}]{2010ApJ...711L..19B}
{Bean} JL, {Seifahrt} A, {Hartman} H, {Nilsson} H, {Reiners} A, {Dreizler} S,
  {Henry} TJ, {Wiedemann} G (2010{\natexlab{a}}) {The Proposed Giant Planet
  Orbiting VB 10 Does Not Exist}. ApJl 711:L19--L23,
  %\doi{10.1088/2041-8205/711/1/L19}

\bibitem[{{Bean} et~al(2010{\natexlab{b}}){Bean}, {Seifahrt}, {Hartman},
  {Nilsson}, {Wiedemann}, {Reiners}, {Dreizler}, and
  {Henry}}]{2010ApJ...713..410B}
{Bean} JL, {Seifahrt} A, {Hartman} H, {Nilsson} H, {Wiedemann} G, {Reiners} A,
  {Dreizler} S, {Henry} TJ (2010{\natexlab{b}}) {The CRIRES Search for Planets
  Around the Lowest-mass Stars. I. High-precision Near-infrared Radial
  Velocities with an Ammonia Gas Cell}. ApJ 713:410--422,
  %\doi{10.1088/0004-637X/713/1/410}, %\eprint{0911.3148}

\bibitem[{{Berta} et~al(2011){Berta}, {Charbonneau}, {Bean}, {Irwin}, {Burke},
  {D{\'e}sert}, {Nutzman}, and {Falco}}]{2011ApJ...736...12B}
{Berta} ZK, {Charbonneau} D, {Bean} J, {Irwin} J, {Burke} CJ, {D{\'e}sert} JM,
  {Nutzman} P, {Falco} EE (2011) {The GJ1214 Super-Earth System: Stellar
  Variability, New Transits, and a Search for Additional Planets}. ApJ 736:12,
  %\doi{10.1088/0004-637X/736/1/12}, %\eprint{1012.0518}

\bibitem[{{Blake} et~al(2010){Blake}, {Charbonneau}, and
  {White}}]{2010ApJ...723..684B}
{Blake} CH, {Charbonneau} D, {White} RJ (2010) {The NIRSPEC Ultracool Dwarf
  Radial Velocity Survey}. ApJ 723:684--706, %\doi{10.1088/0004-637X/723/1/684},
  %\eprint{1008.3874}

\bibitem[{{Charbonneau} et~al(2009){Charbonneau}, {Berta}, {Irwin}, {Burke},
  {Nutzman}, {Buchhave}, {Lovis}, {Bonfils}, {Latham}, {Udry}, {Murray-Clay},
  {Holman}, {Falco}, {Winn}, {Queloz}, {Pepe}, {Mayor}, {Delfosse}, and
  {Forveille}}]{2009Natur.462..891C}
{Charbonneau} D, {Berta} ZK, {Irwin} J, {Burke} CJ, {Nutzman} P, {Buchhave} LA,
  {Lovis} C, {Bonfils} X, {Latham} DW, {Udry} S, {Murray-Clay} RA, {Holman} MJ,
  {Falco} EE, {Winn} JN, {Queloz} D, {Pepe} F, {Mayor} M, {Delfosse} X,
  {Forveille} T (2009) {A super-Earth transiting a nearby low-mass star}. \nat
  462:891--894, %\doi{10.1038/nature08679}, %\eprint{0912.3229}

\bibitem[{{Crifo} et~al(2010){Crifo}, {Jasniewicz}, {Soubiran}, {Katz},
  {Siebert}, {Veltz}, and {Udry}}]{2010A&A...524A..10C}
{Crifo} F, {Jasniewicz} G, {Soubiran} C, {Katz} D, {Siebert} A, {Veltz} L,
  {Udry} S (2010) {Towards a new full-sky list of radial velocity standard
  stars}. \aap 524:A10, %\doi{10.1051/0004-6361/201015315}, %\eprint{1010.0613}

\bibitem[{{Cutri} et~al(2003){Cutri}, {Skrutskie}, {van Dyk}, {Beichman},
  {Carpenter}, {Chester}, {Cambresy}, {Evans}, {Fowler}, {Gizis}, {Howard},
  {Huchra}, {Jarrett}, {Kopan}, {Kirkpatrick}, {Light}, {Marsh}, {McCallon},
  {Schneider}, {Stiening}, {Sykes}, {Weinberg}, {Wheaton}, {Wheelock}, and
  {Zacarias}}]{2003yCat.2246....0C}
{Cutri} RM, {Skrutskie} MF, {van Dyk} S, {Beichman} CA, {Carpenter} JM,
  {Chester} T, {Cambresy} L, {Evans} T, {Fowler} J, {Gizis} J, {Howard} E,
  {Huchra} J, {Jarrett} T, {Kopan} EL, {Kirkpatrick} JD, {Light} RM, {Marsh}
  KA, {McCallon} H, {Schneider} S, {Stiening} R, {Sykes} M, {Weinberg} M,
  {Wheaton} WA, {Wheelock} S, {Zacarias} N (2003) {VizieR Online Data Catalog:
  2MASS All-Sky Catalog of Point Sources (Cutri+ 2003)}. VizieR Online Data
  Catalog 2246

\bibitem[{{Figueira} et~al(2010{\natexlab{a}}){Figueira}, {Pepe}, {Lovis}, and
  {Mayor}}]{2010A&A...515A.106F}
{Figueira} P, {Pepe} F, {Lovis} C, {Mayor} M (2010{\natexlab{a}}) {Evaluating
  the stability of atmospheric lines with HARPS}. \aap 515:A106,
  %\doi{10.1051/0004-6361/201014005}, %\eprint{1003.0541}

\bibitem[{{Figueira} et~al(2010{\natexlab{b}}){Figueira}, {Pepe}, {Melo},
  {Santos}, {Lovis}, {Mayor}, {Queloz}, {Smette}, and
  {Udry}}]{2010A&A...511A..55F}
{Figueira} P, {Pepe} F, {Melo} CHF, {Santos} NC, {Lovis} C, {Mayor} M, {Queloz}
  D, {Smette} A, {Udry} S (2010{\natexlab{b}}) {Radial velocities with CRIRES.
  Pushing precision down to 5-10 m/s}. \aap 511:A55,
  %\doi{10.1051/0004-6361/200912681}, %\eprint{0912.2643}

\bibitem[{{Goldberg} and {M{\"u}ller}(1953)}]{1953ApJ...118..397G}
{Goldberg} L, {M{\"u}ller} EA (1953) {Carbon Monoxide in the Sun.} ApJ 118:397,
  %\doi{10.1086/145768}

\bibitem[{{Henry} et~al(2006){Henry}, {Koerner}, {Jao}, {Subasavage}, {Ianna},
  and {RECONS}}]{2006AAS...209.2404H}
{Henry} TJ, {Koerner} DW, {Jao} WC, {Subasavage} JP, {Ianna} PA, {RECONS}
  (2006) {Clandestine Companions of Nearby Red Dwarfs}. In: American
  Astronomical Society Meeting Abstracts, Bulletin of the American Astronomical
  Society, vol~38, p 933

\bibitem[{{Howard} et~al(2014){Howard}, {Marcy}, {Fischer}, {Isaacson},
  {Muirhead}, {Henry}, {Boyajian}, {von Braun}, {Becker}, {Wright}, and
  {Johnson}}]{2014ApJ...794...51H}
{Howard} AW, {Marcy} GW, {Fischer} DA, {Isaacson} H, {Muirhead} PS, {Henry} GW,
  {Boyajian} TS, {von Braun} K, {Becker} JC, {Wright} JT, {Johnson} JA (2014)
  {The NASA-UC-UH ETA-Earth Program. IV. A Low-mass Planet Orbiting an M Dwarf
  3.6 PC from Earth}. ApJ 794:51, %\doi{10.1088/0004-637X/794/1/51},
  %\eprint{1408.5645}

\bibitem[{{Hu{\'e}lamo} et~al(2008){Hu{\'e}lamo}, {Figueira}, {Bonfils},
  {Santos}, {Pepe}, {Gillon}, {Azevedo}, {Barman}, {Fern{\'a}ndez}, {di Folco},
  {Guenther}, {Lovis}, {Melo}, {Queloz}, and {Udry}}]{2008A&A...489L...9H}
{Hu{\'e}lamo} N, {Figueira} P, {Bonfils} X, {Santos} NC, {Pepe} F, {Gillon} M,
  {Azevedo} R, {Barman} T, {Fern{\'a}ndez} M, {di Folco} E, {Guenther} EW,
  {Lovis} C, {Melo} CHF, {Queloz} D, {Udry} S (2008) {TW Hydrae: evidence of
  stellar spots instead of a Hot Jupiter}. \aap 489:L9--L13,
  %\doi{10.1051/0004-6361:200810596}, %\eprint{0808.2386}

\bibitem[{{Isaacson} and {Fischer}(2010)}]{2010ApJ...725..875I}
{Isaacson} H, {Fischer} D (2010) {Chromospheric Activity and Jitter
  Measurements for 2630 Stars on the California Planet Search}. ApJ
  725:875--885, %\doi{10.1088/0004-637X/725/1/875}, %\eprint{1009.2301}

\bibitem[{{Kaltenegger} and {Traub}(2009)}]{2009ApJ...698..519K}
{Kaltenegger} L, {Traub} WA (2009) {Transits of Earth-like Planets}. ApJ
  698:519--527, %\doi{10.1088/0004-637X/698/1/519}, %\eprint{0903.3371}

\bibitem[{{Kasting} et~al(1993){Kasting}, {Whitmire}, and
  {Reynolds}}]{1993Icar..101..108K}
{Kasting} JF, {Whitmire} DP, {Reynolds} RT (1993) {Habitable Zones around Main
  Sequence Stars}. \icarus 101:108--128, %\doi{10.1006/icar.1993.1010}

\bibitem[{{L{\'e}pine} and {Gaidos}(2011)}]{2011AJ....142..138L}
{L{\'e}pine} S, {Gaidos} E (2011) {An All-sky Catalog of Bright M Dwarfs}. \aj
  142:138, %\doi{10.1088/0004-6256/142/4/138}, %\eprint{1108.2719}

\bibitem[{{Lovis} et~al(2006){Lovis}, {Mayor}, {Pepe}, {Alibert}, {Benz},
  {Bouchy}, {Correia}, {Laskar}, {Mordasini}, {Queloz}, {Santos}, {Udry},
  {Bertaux}, and {Sivan}}]{2006Natur.441..305L}
{Lovis} C, {Mayor} M, {Pepe} F, {Alibert} Y, {Benz} W, {Bouchy} F, {Correia}
  ACM, {Laskar} J, {Mordasini} C, {Queloz} D, {Santos} NC, {Udry} S, {Bertaux}
  JL, {Sivan} JP (2006) {An extrasolar planetary system with three Neptune-mass
  planets}. \nat 441:305--309, %\doi{10.1038/nature04828},
  %\eprint{astro-ph/0703024}

\bibitem[{{Mart{\'{\i}}nez-Arn{\'a}iz} et~al(2010){Mart{\'{\i}}nez-Arn{\'a}iz},
  {Maldonado}, {Montes}, {Eiroa}, and {Montesinos}}]{2010A&A...520A..79M}
{Mart{\'{\i}}nez-Arn{\'a}iz} R, {Maldonado} J, {Montes} D, {Eiroa} C,
  {Montesinos} B (2010) {Chromospheric activity and rotation of FGK stars in
  the solar vicinity. An estimation of the radial velocity jitter}. \aap
  520:A79, %\doi{10.1051/0004-6361/200913725}, %\eprint{1002.4391}

\bibitem[{{Mayor} et~al(1995){Mayor}, {Queloz}, {Marcy}, {Butler}, {Noyes},
  {Korzennik}, {Krockenberger}, {Nisenson}, {Brown}, {Kennelly}, {Rowland},
  {Horner}, {Burki}, {Burnet}, and {Kunzli}}]{1995IAUC.6251....1M}
{Mayor} M, {Queloz} D, {Marcy} G, {Butler} P, {Noyes} R, {Korzennik} S,
  {Krockenberger} M, {Nisenson} P, {Brown} T, {Kennelly} T, {Rowland} C,
  {Horner} S, {Burki} G, {Burnet} M, {Kunzli} M (1995) {51 Pegasi}. \iaucirc
  6251

\bibitem[{{Mishenina} et~al(2004){Mishenina}, {Soubiran}, {Kovtyukh}, and
  {Korotin}}]{2004A&A...418..551M}
{Mishenina} TV, {Soubiran} C, {Kovtyukh} VV, {Korotin} SA (2004) {On the
  correlation of elemental abundances with kinematics among galactic disk
  stars}. \aap 418:551--562, %\doi{10.1051/0004-6361:20034454},
  %\eprint{astro-ph/0401234}

\bibitem[{{Mohanty} and {Basri}(2003)}]{2003ApJ...583..451M}
{Mohanty} S, {Basri} G (2003) {Rotation and Activity in Mid-M to L Field
  Dwarfs}. ApJ 583:451--472, %\doi{10.1086/345097}, %\eprint{astro-ph/0201455}

\bibitem[{{Mohler} et~al(1953){Mohler}, {Pierce}, {McMath}, and
  {Goldberg}}]{1953ApJ...117...41M}
{Mohler} OC, {Pierce} AK, {McMath} RR, {Goldberg} L (1953) {Table of Infrared
  Solar Lines, 1.4-2.5 {$\mu$}.} ApJ 117:41, %\doi{10.1086/145667}

\bibitem[{{Oliva} et~al(2006){Oliva}, {Origlia}, {Baffa}, {Biliotti}, {Bruno},
  {D'Amato}, {Del Vecchio}, {Falcini}, {Gennari}, {Ghinassi}, {Giani},
  {Gonzalez}, {Leone}, {Lolli}, {Lodi}, {Maiolino}, {Mannucci}, {Marcucci},
  {Mochi}, {Montegriffo}, {Rossetti}, {Scuderi}, and
  {Sozzi}}]{2006SPIE.6269E..19O}
{Oliva} E, {Origlia} L, {Baffa} C, {Biliotti} C, {Bruno} P, {D'Amato} F, {Del
  Vecchio} C, {Falcini} G, {Gennari} S, {Ghinassi} F, {Giani} E, {Gonzalez} M,
  {Leone} F, {Lolli} M, {Lodi} M, {Maiolino} R, {Mannucci} F, {Marcucci} G,
  {Mochi} I, {Montegriffo} P, {Rossetti} E, {Scuderi} S, {Sozzi} M (2006) {The
  GIANO-TNG spectrometer}. In: Society of Photo-Optical Instrumentation
  Engineers (SPIE) Conference Series, \procspie, vol 6269, p 626919,
  %\doi{10.1117/12.670006}

\bibitem[{{Oliva} et~al(2012){Oliva}, {Origlia}, {Maiolino}, {Baffa},
  {Biliotti}, {Bruno}, {Falcini}, {Gavriousev}, {Ghinassi}, {Giani},
  {Gonzalez}, {Leone}, {Lodi}, {Massi}, {Mochi}, {Montegriffo}, {Pedani},
  {Rossetti}, {Scuderi}, {Sozzi}, and {Tozzi}}]{2012SPIE.8446E..3TO}
{Oliva} E, {Origlia} L, {Maiolino} R, {Baffa} C, {Biliotti} V, {Bruno} P,
  {Falcini} G, {Gavriousev} V, {Ghinassi} F, {Giani} E, {Gonzalez} M, {Leone}
  F, {Lodi} M, {Massi} F, {Mochi} I, {Montegriffo} P, {Pedani} M, {Rossetti} E,
  {Scuderi} S, {Sozzi} M, {Tozzi} A (2012) {The GIANO spectrometer: towards its
  first light at the TNG}. In: Ground-based and Airborne Instrumentation for
  Astronomy IV, \procspie, vol 8446, p 84463T, %\doi{10.1117/12.925274}

\bibitem[{{Origlia} et~al(2014){Origlia}, {Oliva}, {Baffa}, {Falcini}, {Giani},
  {Massi}, {Montegriffo}, {Sanna}, {Scuderi}, {Sozzi}, {Tozzi}, {Carleo},
  {Gratton}, {Ghinassi}, and {Lodi}}]{2014SPIE.9147E..1EO}
{Origlia} L, {Oliva} E, {Baffa} C, {Falcini} G, {Giani} E, {Massi} F,
  {Montegriffo} P, {Sanna} N, {Scuderi} S, {Sozzi} M, {Tozzi} A, {Carleo} I,
  {Gratton} R, {Ghinassi} F, {Lodi} M (2014) {High resolution near IR
  spectroscopy with GIANO-TNG}. In: Ground-based and Airborne Instrumentation
  for Astronomy V, \procspie, vol 9147, p 91471E, %\doi{10.1117/12.2054743}

\bibitem[{{Pepe} et~al(2004){Pepe}, {Mayor}, {Queloz}, {Benz}, {Bonfils},
  {Bouchy}, {Lo Curto}, {Lovis}, {M{\'e}gevand}, {Moutou}, {Naef}, {Rupprecht},
  {Santos}, {Sivan}, {Sosnowska}, and {Udry}}]{2004A&A...423..385P}
{Pepe} F, {Mayor} M, {Queloz} D, {Benz} W, {Bonfils} X, {Bouchy} F, {Lo Curto}
  G, {Lovis} C, {M{\'e}gevand} D, {Moutou} C, {Naef} D, {Rupprecht} G, {Santos}
  NC, {Sivan} JP, {Sosnowska} D, {Udry} S (2004) {The HARPS search for southern
  extra-solar planets. I. HD 330075 b: A new ``hot Jupiter''}. \aap
  423:385--389, %\doi{10.1051/0004-6361:20040389}, %\eprint{astro-ph/0405252}

\bibitem[{{Pravdo} and {Shaklan}(2009)}]{2009ApJ...700..623P}
{Pravdo} SH, {Shaklan} SB (2009) {An ultracool Star's Candidate Planet}. ApJ
  700:623--632, %\doi{10.1088/0004-637X/700/1/623}, %\eprint{0906.0544}

\bibitem[{{Queloz} et~al(2001){Queloz}, {Henry}, {Sivan}, {Baliunas}, {Beuzit},
  {Donahue}, {Mayor}, {Naef}, {Perrier}, and {Udry}}]{2001A&A...379..279Q}
{Queloz} D, {Henry} GW, {Sivan} JP, {Baliunas} SL, {Beuzit} JL, {Donahue} RA,
  {Mayor} M, {Naef} D, {Perrier} C, {Udry} S (2001) {No planet for HD 166435}.
  \aap 379:279--287, %\doi{10.1051/0004-6361:20011308},
  %\eprint{astro-ph/0109491}

\bibitem[{{Rodler} et~al(2012){Rodler}, {Deshpande}, {Zapatero Osorio},
  {Mart{\'{\i}}n}, {Montgomery}, {Del Burgo}, and
  {Creevey}}]{2012A&A...538A.141R}
{Rodler} F, {Deshpande} R, {Zapatero Osorio} MR, {Mart{\'{\i}}n} EL,
  {Montgomery} MM, {Del Burgo} C, {Creevey} OL (2012) {Search for radial
  velocity variations in eight M-dwarfs with NIRSPEC/Keck II}. \aap 538:A141,
  %\doi{10.1051/0004-6361/201117577}, %\eprint{1112.1382}

\bibitem[{{Seifahrt} and {K{\"a}ufl}(2008)}]{2008A&A...491..929S}
{Seifahrt} A, {K{\"a}ufl} HU (2008) {High precision radial velocity
  measurements in the infrared. A first assessment of the RV stability of
  CRIRES}. \aap 491:929--939, %\doi{10.1051/0004-6361:200810174},
  %\eprint{0809.1435}

\bibitem[{{Setiawan} et~al(2008){Setiawan}, {Henning}, {Launhardt},
  {M{\"u}ller}, {Weise}, and {K{\"u}rster}}]{2008Natur.451...38S}
{Setiawan} J, {Henning} T, {Launhardt} R, {M{\"u}ller} A, {Weise} P,
  {K{\"u}rster} M (2008) {A young massive planet in a star-disk system}. \nat
  451:38--41, %\doi{10.1038/nature06426}

\bibitem[{{Strassmeier} et~al(2000){Strassmeier}, {Washuettl}, {Granzer},
  {Scheck}, and {Weber}}]{2000A&AS..142..275S}
{Strassmeier} K, {Washuettl} A, {Granzer} T, {Scheck} M, {Weber} M (2000) {The
  Vienna-KPNO search for Doppler-imaging candidate stars. I. A catalog of
  stellar-activity indicators for 1058 late-type Hipparcos stars}. \aaps
  142:275--311, %\doi{10.1051/aas:2000328}

\bibitem[{{Tozzi} et~al(2014){Tozzi}, {Oliva}, {Origlia}, {Baffa}, {Biliotti},
  {Falcini}, {Giani}, {Iuzzolino}, {Massi}, {Sanna}, {Scuderi}, and
  {Sozzi}}]{2014SPIE.9147E..9NT}
{Tozzi} A, {Oliva} E, {Origlia} L, {Baffa} C, {Biliotti} V, {Falcini} G,
  {Giani} E, {Iuzzolino} M, {Massi} F, {Sanna} N, {Scuderi} S, {Sozzi} M (2014)
  {The fiber-fed preslit of GIANO at T.N.G.} In: Ground-based and Airborne
  Instrumentation for Astronomy V, \procspie, vol 9147, p 91479N,
  %\doi{10.1117/12.2054094}, %\eprint{1407.3126}


\end{thebibliography}

\begin{appendix}

\begin{table*}[tb]
\centering
\caption{Final results for HD3765. For each observation we have RV value, its internal error, the bisector velocity span of both telluric and star.}
\begin{tabular}{clccccc}
\toprule
Exp. Number & JD-2450000 & Original RV & RV after correction & Final Error  & $BVS_{star}$  & $BVS_{tell}$  \\
 &  &  (km\, s$^{-1}$) & (km\, s$^{-1}$) & (km\, s$^{-1}$) &  (km\, s$^{-1}$) & (km\, s$^{-1}$) \\
\midrule
0  & 6907.56615741 & -63.2389 & -63.2616  & 0.0047   & -0.1651  &-0.1651 \\
1  & 6907.57447917 & -63.2461 & -63.2697  & 0.0055   & -0.1618  &-0.0659 \\
2  & 6907.58274306 & -63.2396 & -63.2628  & 0.0058   & -0.1636  &-0.0660 \\
3  & 6907.59107639 & -63.2402 & -63.2580  & 0.0063   & -0.1723  &-0.0531 \\
4  & 6910.43225694 & -63.3021 & -63.2572  & 0.0075   & -0.2628  & 0.1109 \\
5  & 6910.44055556 & -63.2855 & -63.2426  & 0.0068   & -0.2636  & 0.1020 \\
6  & 6910.44886574 & -63.3037 & -63.2595  & 0.0069   & -0.2570  & 0.1141 \\
7  & 6911.69504630 & -63.2726 & -63.2874  & 0.0063   & -0.1201  & 0.0113 \\
8  & 6911.70334491 & -63.2475 & -63.2594  & 0.0058   & -0.1330  & 0.0104 \\
9  & 6911.71164352 & -63.2533 & -63.2644  & 0.0057   & -0.1302  & 0.0163 \\
10 & 6911.71994213 & -63.2435 & -63.2557  & 0.0055   & -0.1312  & 0.0107 \\
11 & 6912.65265046 & -63.2492 & -63.2565  & 0.0068   & -0.1378  & 0.0240 \\
12 & 6912.66100694 & -63.2481 & -63.2520  & 0.0068   & -0.1492  & 0.0267 \\
13 & 6912.66934028 & -63.2501 & -63.2556  & 0.0067   & -0.1517  & 0.0177 \\
14 & 6913.65021991 & -63.2353 & -63.2529  & 0.0077   & -0.1013  & 0.0188 \\
15 & 6913.65861111 & -63.2230 & -63.2422  & 0.0071   & -0.0992  & 0.0144 \\
16 & 6913.66699074 & -63.2328 & -63.2506  & 0.0070   & -0.0996  & 0.0197 \\
17 & 6914.65011574 & -63.2495 & -63.2555  & 0.0104   & -0.0940  & 0.0734 \\
18 & 6914.65855324 & -63.1921 & -63.2098  & 0.0104   & -0.0539  & 0.0659 \\
19 & 6914.66695602 & -63.2510 & -63.2696  & 0.0075   & -0.0742  & 0.0417 \\
20 & 6916.58828704 & -63.2962 & -63.2693  & 0.0082   & -0.2048  & 0.0959 \\
21 & 6916.59723380 & -63.2965 & -63.2722  & 0.0086   & -0.1934  & 0.0969 \\
22 & 6916.60608796 & -63.2902 & -63.2629  & 0.0079   & -0.1989  & 0.1033 \\

\bottomrule
\end{tabular}
\label{tab:hd3765_0}
\end{table*}

\begin{table*}[tb]
\centering
\caption{Final results for GJ1214. For each observation we have RV value, its internal error, the bisector velocity span of both telluric and star.}
\begin{tabular}{clccccc}
\toprule
Exp. Number & JD-2450000 & Original RV & RV after correction  & Final Error  & $BVS_{star}$  & $BVS_{tell}$  \\
 &  &  (km\, s$^{-1}$) & (km\, s$^{-1}$) & (km\, s$^{-1}$) &  (km\, s$^{-1}$) & (km\, s$^{-1}$) \\
\midrule
0 & 6910.38541667 & 20.9418 & 21.0369 & 0.0369   &  0.1442 &  0.0934 \\
1 & 6910.38541667 & 20.8088 & 20.9044 & 0.0418   & -0.0667 &  0.1521 \\
2 & 6910.40211806 & 21.0155 & 21.1029 & 0.0430   & -0.0668 &  0.0487 \\
3 & 6910.41050926 & 21.023  & 21.1019 & 0.0475   &  0.1265 &  0.1160 \\
4 & 6911.38696759 & 21.2712  & 21.1724 & 0.0339   &  0.1718 &  0.1249 \\
5 & 6911.39531250 & 21.3144  & 21.1956 & 0.0360   &  0.1176 &  0.1097 \\
6 & 6911.40375000 & 21.3013 & 21.1692 & 0.0383   & -0.0742 & -0.0434 \\
7 & 6911.41210648 & 21.0979 & 20.9203 & 0.0386   &  0.0106 &  0.0160 \\
8 & 6912.36031250 & 21.1599 & 21.0876 & 0.0309   &  0.0996 &  0.0654 \\
9 & 6912.36866898 & 21.1201 & 21.0445 & 0.0289   & -0.0393 &  0.0463 \\
10& 6912.37703704 & 21.0178 & 20.9127 & 0.0304   &  0.0144 &  0.0288 \\
11& 6912.38539352 & 20.942  & 20.7980 & 0.0319   & -0.0123 &  0.0988 \\
12& 6913.35280093 & 20.9556 & 20.9571 & 0.0287   &  0.1305 &  0.0561 \\
13& 6913.36111111 & 21.1827 & 21.1572 & 0.0328   &  0.0553 &  0.0635 \\
14& 6913.36952546 & 21.0057 & 20.9675 & 0.0314   & -0.0392 &  0.0006 \\
15& 6913.37789352 & 21.0947 & 21.0550 & 0.0308   &  0.0873 &  0.0005 \\
16& 6914.34678241 & 20.9316 & 20.9721 & 0.0322   & -0.0621 &  0.0179 \\
17& 6914.35511574 & 20.8875 & 20.9107 & 0.0318   & -0.0161 & -0.0427 \\
18& 6914.36344907 & 20.9097 & 20.9081 & 0.0306   & -0.0363 & -0.0199 \\
19& 6914.37181713 & 21.0563 & 21.0343 & 0.0337   & -0.0311 &  0.0279 \\

\bottomrule
\end{tabular}
\label{tab:gj1214_results}
\end{table*}

\begin{table*}[tb]
\centering
\caption{Final results for Gl15A. For each observation we have RV value, its internal error, the bisector velocity span of both telluric and star.}
\begin{tabular}{clccccc}
\toprule
Exp. Number & JD-2450000 & Original RV & RV after correction & Final Error  & $BVS_{star}$  & $BVS_{tell}$  \\
 &  &  (km\, s$^{-1}$) &  (km\, s$^{-1}$) & (km\, s$^{-1}$) & (km\, s$^{-1}$) & (km\, s$^{-1}$) \\
\midrule
 0 & 6907.60567130 & 11.6598 & 11.6208 & 0.0031   & -0.0662  &-0.0205 \\
 1 & 6907.61401620 & 11.6629 & 11.6243 & 0.0028   & -0.0720  &-0.0216 \\
 2 & 6907.62231481 & 11.6680 & 11.6315 & 0.0027   & -0.0731  &-0.0196 \\
 3 & 6907.63057870 & 11.6772 & 11.6416 & 0.0027   & -0.0645  &-0.0206 \\
 4 & 6911.65913194 & 11.5881 & 11.6212 & 0.0036   & -0.0376  & 0.0318 \\
 5 & 6911.66746528 & 11.5834 & 11.6190 & 0.0036   & -0.0323  & 0.0361 \\
 6 & 6911.67576389 & 11.5786 & 11.6170 & 0.0037   & -0.0387  & 0.0370 \\
 7 & 6911.68406250 & 11.5980 & 11.6396 & 0.0036   & -0.0302  & 0.0336 \\
 8 & 6912.62534722 & 11.6051 & 11.6186 & 0.0042   & -0.0022  & 0.0360 \\
 9 & 6912.63368056 & 11.6136 & 11.6258 & 0.0037   & -0.0089  & 0.0458 \\
10 & 6912.64207176 & 11.6182 & 11.6306 & 0.0034   & -0.0169  & 0.0519 \\
11 & 6913.62258102 & 11.6145 & 11.6330 & 0.0044   & -0.0280  & 0.0626 \\
12 & 6913.63091435 & 11.6445 & 11.6615 & 0.0044   & -0.0127  & 0.0653 \\
13 & 6913.63931713 & 11.6329 & 11.6519 & 0.0041   & -0.0325  & 0.0688 \\
14 & 6914.62081019 & 11.6172 & 11.6162 & 0.0047   & -0.0694  & 0.0657 \\
15 & 6914.62913194 & 11.5912 & 11.5912 & 0.0045   & -0.0804  & 0.0612 \\
16 & 6914.63744213 & 11.5727 & 11.5756 & 0.0050   & -0.0888  & 0.0771 \\
17 & 6916.55939815 & 11.6580 & 11.6282 & 0.0051   & -0.0573  & 0.1247 \\
18 & 6916.56820602 & 11.6654 & 11.6352 & 0.0054   & -0.0531  & 0.1216 \\
19 & 6916.57704861 & 11.6623 & 11.6317 & 0.0056   & -0.0390  & 0.1150 \\

\bottomrule
\end{tabular}
\label{tab:gl15a_results}
\end{table*}

\begin{table*}[tb]
\centering
\caption{Final results for VB10. For each observation we have RV value, its internal error, the bisector velocity span of both telluric and star.}
\begin{tabular}{clccccc}
\toprule
Exp. Number & JD-2450000 & Original RV &  RV after correction & Final Error  & $BVS_{star}$  & $BVS_{tell}$  \\
 &  &  (km\, s$^{-1}$) &  (km\, s$^{-1}$) & (km\, s$^{-1}$) & (km\, s$^{-1}$) & (km\, s$^{-1}$) \\
\midrule
 0 & 6911.46657407 & -35.6947 & -35.6979 & 0.0343  & -0.1512  & 0.0558 \\
 1 & 6911.47489583 & -35.8025 & -35.7180 & 0.0323  & -0.0843  & 0.0105 \\
 2 & 6911.48322917 & -35.6241 & -35.5388 & 0.0325  & -0.1073  & 0.1207 \\
 3 & 6911.51429398 & -35.8705 & -35.5881 & 0.0361  & -0.1519  & 0.1976 \\
 4 & 6911.52851852 & -35.5356 & -35.1693 & 0.0361  & -0.0798  & 0.2147 \\
 5 & 6912.47067130 & -35.4708 & -35.4721 & 0.0301  &  0.0033  & 0.1100 \\
 6 & 6912.47901620 & -35.7099 & -35.6565 & 0.0295  & -0.0695  & 0.2161 \\
 7 & 6912.48747685 & -35.5564 & -35.4703 & 0.0274  &  0.0275  & 0.1721 \\
 8 & 6912.49581019 & -35.7067 & -35.5660 & 0.0270  & -0.0277  & 0.2216 \\
 9 & 6913.45810185 & -35.5303 & -35.6098 & 0.0285  & -0.0712  & 0.1600 \\
10 & 6913.46645833 & -35.5952 & -35.6481 & 0.0266  & -0.1164  & 0.2621 \\
11 & 6913.47489583 & -35.6751 & -35.7056 & 0.0271  &  0.0148  & 0.3159 \\
12 & 6913.48322917 & -35.5411 & -35.4999 & 0.0254  & -0.0522  & 0.3453 \\
13 & 6914.44444444 & -35.5564 & -35.6165 & 0.0275  &  0.0904  & 0.2588 \\
14 & 6914.45287037 & -35.3866 & -35.4494 & 0.0362  &  0.0635  & 0.3648 \\
15 & 6914.46130787 & -35.4753 & -35.5016 & 0.0291  &  0.0354  & 0.2125 \\
16 & 6914.46974537 & -35.4767 & -35.4814 & 0.0263  & -0.0652  & 0.3792 \\

\bottomrule
\end{tabular}
\label{tab:vb10_results}
\end{table*}

\end{appendix}

\end{document}